\documentclass[10pt]{iopart}
\pdfoutput=1
\usepackage{graphicx}
\usepackage{bm}
\usepackage{cite}

\newcommand{\avec}[1]{{\bm{#1}}}
\newcommand{\avecu}[1]{{\hat{\bm{#1}}}}

\begin{document}

\title{Molecular dynamics simulation: a tool for exploration and discovery
using simple models}

\author{D.C. Rapaport} 

\address{Department of Physics, Bar-Ilan University, Ramat-Gan, Israel 52900}
\ead{rapaport@mail.biu.ac.il}

\vspace{10pt}

\begin{indented}
\item[]22 October 2014
\end{indented}

\begin{abstract}

Emergent phenomena share the fascinating property of not being obvious
consequences of the design of the system in which they appear. This
characteristic is no less relevant when attempting to simulate such
phenomena, given that the outcome is not always a foregone conclusion. The
present survey focuses on several simple model systems that exhibit
surprisingly rich emergent behavior, all studied by molecular dynamics (MD)
simulation. The examples are taken from the disparate fields of fluid
dynamics, granular matter and supramolecular self-assembly. In studies of
fluids modeled at the detailed microscopic level using discrete particles,
the simulations demonstrate that complex hydrodynamic phenomena in rotating
and convecting fluids -- the Taylor--Couette and Rayleigh--B\'enard
instabilities -- can not only be observed within the limited length and time
scales accessible to MD, but even quantitative agreement can be achieved.
Simulation of highly counterintuitive segregation phenomena in granular
mixtures, again using MD methods, but now augmented by forces producing
damping and friction, leads to results that resemble experimentally observed
axial and radial segregation in the case of a rotating cylinder, and to a
novel form of horizontal segregation in a vertically vibrated layer.
Finally, when modeling self-assembly processes analogous to the formation of
the polyhedral shells that package spherical viruses, simulation of suitably
shaped particles reveals the ability to produce complete, error-free
assembly, and leads to the important general observation that reversible
growth steps contribute to the high yield. While there are limitations to
the MD approach, both computational and conceptual, the results offer a
tantalizing hint of the kinds of phenomena that can be explored, and what
might be discovered when sufficient resources are brought to bear on a
problem.

\end{abstract}

\vspace{2pc}


\ioptwocol

\section{Introduction}

The behavior associated with emergent phenomena is unusual in the sense that
it is not an obvious consequence of the system design. `Emergence', where
the whole appears to be more than just the sum of its parts, is a term in
widespread use, particularly in the biological and social sciences. However,
when used to characterize effects that are inherently physical, the presence
of some unknown component is not implied, but merely a manifestation of
cooperative behavior that the theoretician's tools are unable to deduce from
the underlying equations. (There is a certain arbitrariness in designating
phenomena as emergent, with the field of phase transitions often excluded,
probably due to its theoretical foundations predating the popularization of
the concept).

Computer simulation provides an ideal framework for studying physical
phenomena that exhibit emergent characteristics. Simulation, unlike mere
computation, aims to accommodate a more flexible outcome, thereby
facilitating `exploration'; if the unexpected is encountered, this amounts
to `discovery'. The discovery of emergent behavior in the course of a
simulation is arguably even more surprising than in the real world, since
the ingredients of the model system are (by definition) known, whereas in
nature, as a last resort, emergence can always be attributed to mechanisms
not fully understood.

The systems described in this article all exhibit emergent behavior. They
are presented as a series of case studies and are based, for obvious
reasons, on the author's own work. The case studies, all involving classical
molecular dynamics (MD) simulation, fall into distinct categories and are
described in greater detail in subsequent sections of the paper; this
grouping reflects the ability of emergence to manifest itself in different
ways. The problems investigated can be summarized as follows: (a) Complex
hydrodynamic behavior in atomistic fluids: the Taylor--Couette and
Rayleigh--B\'enard instabilities where structured flow patterns develop
under the appropriate conditions. (b) Granular segregation: the components
of a binary mixture of particles with different properties undergo axial
and/or radial segregation in a rotating cylinder, or horizontal segregation
in a layer placed on a vertically vibrated base with sawtooth grooves. (c)
Supramolecular self-assembly: the spontaneous formation of polyhedral shells
from particles of suitable shape, behavior analogous to the growth of
protein shells that encapsulate spherical viruses.

Emergent behavior is associated with effects occurring at a `higher' level
than the participating elements. In the case of complex fluid flow, the
atoms have little `awareness' of collective movement on spatial and temporal
scales far beyond the mean interatomic separation and the mean time between
collisions. The same is true for segregating grains, although most will
gradually become aware that they are surrounded by similar neighbors. With
self-assembly, the molecular constituents are continually forming and
breaking bonds, but gradually find themselves completely bonded, with little
or no possibility of escape, as they are incorporated into the growing
shells. In each of the examples, individual particles respond to the forces
exerted by their immediate surroundings and, where relevant, boundary walls
and an external force such as gravity; the common characteristic shared by
these systems is that, if and when large-scale coherent behavior does
emerge, it does so spontaneously.

Although the systems described here are mostly familiar, the {\em a priori}
expectation of what simulations might be capable of achieving range from the
uninteresting case of nothing, to the ideal outcome in which the rich
complexity  of the real world is reproduced. In studies of fluid flow, the
surprise would be that macroscopic phenomena actually persist down to the
microscopic scales accessible to MD. A successful outcome may show
qualitatively meaningful behavior or even quantitative agreement with
experiment and/or theory. Failure could reflect inadequacy of the model,
unsuitable parameter choices, or simulations of systems that are too small
to accommodate the range of length scales involved or run for insufficient
time for the phenomena to develop. Resolution of the size and time
limitations is aided to a modest extent by advances in computer performance.
Spurious artifacts, implying behavior not observed in nature, are also a
possibility. It goes without saying that mere reproduction of known
behavior, no matter how complex, is not the ultimate goal; once a phenomenon
has been captured {\em `in silico'}, questions can be asked concerning
mechanisms that might be experimentally inaccessible, with a view to
investigating the cooperativity ultimately responsible for emergent
behavior.

Direct observation is crucial for the study of complex phenomena in the real
world, the reason being that quantitative characterization is not always
possible, and even when it is, the information supplied may be incapable of
describing the intricate details. Computer visualization plays the
corresponding role in the simulational environment. Actual observation of
complicated and sometimes unexpected behavior arising in a simulation is
also an illuminating experience because it reinforces the appreciation that
complexity can have simple origins. The results included here emphasize the
visual aspect, and are supplemented with limited quantitative details when
theoretical or experimental comparison is possible or if the description of
the phenomenon is enhanced by numerical data; further analysis is deferred
to subsequent publications that focus on the individual systems. 

The case studies are all based on new simulations that revisit problems
considered in the past. The updated computations take advantage of recent
advances in MD methodology enabled by GPU-based (GPU = graphic processing
unit) computing that offer a 1--2 orders of magnitude performance
improvement (depending on the type of computation and when the earlier work
was carried out) to increase the system size and/or time covered. Larger
systems can reduce finite-size effects, and in some instances, described
later, the models themselves have been enhanced in various ways. The effect
of these changes is reflected in the results. Improved computing power
increases the scope for exploration; some of the examples considered here
that originally required weeks of computing can now be completed in a single
day, those needing months now complete in a week, while others of a less
demanding (but still substantial) nature can now even be carried out
interactively; the implications for proposed future study of these and
related systems are obvious.

The organization of the paper is as follows. General aspects of MD
simulation methodology are briefly summarized in Section~2, including model
design and parameterization, algorithms and optimization for modern
high-performance processors, inherent limitations of the MD approach due to
the simplified models and heavy resource requirements, and the issue of
reproducibility that arises when dealing with emergent behavior. The
individual case studies are described in Sections~3--5: the background to
each of the problems is covered, and the customized models and extensions to
general MD procedures required for the simulations introduced, together with
their limitations; the results accompanying the studies emphasize the more
unexpected aspects of the behavior. Finally, Section~6 provides a
retrospective outlook, attempting to strike a balance between the
capabilities of the simulational approach and the rewards anticipated from
the exploration of problems of increasing size and complexity.

\section{Methods}

Molecular dynamics simulation, or MD, a well-established approach conceived
early in the computer age \cite{ald57,rah64}, is aimed at probing the
detailed dynamics of an extremely broad range of atomic and molecular
systems, leading, among other notable accomplishments, to an understanding
of liquid structure and dynamics, and helping gain a recent (2013) Nobel
Prize for contributions to biomolecular dynamics. The methodology, e.g.,
\cite{rap04bk}, is based on classical dynamics, and while quantum theory
often underlies the models, in particular the interaction potentials,
explicit quantum effects are usually excluded. Simplified models are a key
requirement to avoid being overwhelmed by excessive molecular detail;
however, while it is essential to identify the important elements, it is not
always obvious what assumptions are allowed. As with any theory,
confirmation of whether the outcome of a simulation mirrors reality, either
qualitatively or quantitatively, or is a mere artifice of the model, must
come from outside the simulational framework.

While each of the case studies involves discrete particles and MD
methodology, particles can represent different kinds of objects. For the
fluid studies, unlike conventional computational and theoretical fluid
dynamics that regard a fluid as a continuous medium, the MD approach adopts
an atomistic viewpoint, using the simplest possible soft-sphere particles to
represent the individual atoms or molecules; intrinsic properties of the
fluid, such as the viscosity, thermal expansion coefficient and heat
capacity, are determined by the system itself rather than specified as
parameters. When modeling granular matter, the model particles correspond to
grains; granular simulations generally consider spherical particles, using
dissipative, velocity-dependent interactions as well as history-dependent
interactions to emulate the complexities of friction, but in one of the
current studies the particles consist of rigid tetrahedral arrays of soft
spheres allowing the frictional forces to be simplified (partly for
computational reasons). Finally, in the case of self-assembly, each protein
capsomer (as the assembling particles are known) is again represented by a
set of soft spheres arranged on a suitable rigid framework, with attractive
forces acting selectively between interaction sites located on different
particles to produce bonding; the particles themselves are immersed in a
solvent consisting of soft-sphere atoms. 

Each of the systems considered involves excluded-volume interactions between
the soft spheres; there are additional interactions, treated in later
sections, that are problem specific. Two versions of the soft-sphere
interaction are used. For the fluid and self-assembly simulations, the force
is based on the truncated Lennard-Jones potential
\begin{equation}
u_s(r) = 4 \epsilon \, [(\sigma / r)^{12} - (\sigma / r)^{6}]
  \qquad r < r_c \label{eq:ssint}
\end{equation}
where $r = |\avec{r}|$ is the distance between particle centers; the cutoff
range, $r_c = 2^{1/6} \sigma$, ensures that the force is repulsive. In the
granular simulations, the force depends linearly on particle overlap and is
derived from the potential
\begin{equation}
u_\ell(r) = (k_n / 2) \, (r_c - r)^2 \qquad r < r_c \label{eq:ovint}
\end{equation}
where $k_n$ is the force constant, and the cutoff range is determined by the
mean diameter of the particles involved. In both cases the interactions
ensure that the maximum particle overlap (which reflects the effective
particle diameter and depends on the relative speed at impact and on local
stresses) is minimal.

The use of reduced (dimensionless) MD units is implicit; though not required
here, results expressed in reduced units are readily converted to physical
units for comparison with experiment. For simulations of atoms or molecules,
the reduced unit of length is expressed in terms of $\sigma$, which for
argon is 3.4\,\AA\ (a typical value), and spheres have unit mass; the time
unit is determined by $\epsilon$ and corresponds to $2.16 \times
10^{-12}$\,s (also for argon). Setting the Boltzmann constant to unity
defines the temperature unit. For the granular simulations, length is
defined in terms of the nominal particle diameter (if there is more than one
species the smallest size is used), and to maintain consistency with earlier
work $r_c$ is set to $2^{1/6} \times$ (mean diameter); the diameter is
typically $10^{-3}$\,m, and the time unit is then determined from the value
of the gravitational acceleration $g$ so that the rotation or vibration
rates correspond to experiment (described later). All the simulations use an
integration timestep of 0.005 (MD units).  

General MD methodology covers a variety of algorithms and computational
tasks required by the simulations, including the organization of the force
evaluations in a manner that scales linearly with the number of particles,
the integration of the equations of motion for long simulations while
ensuring stability, rigid-body dynamics, treatment of boundary conditions
and the system initialization process, as well as details specific to
individual problems including analysis techniques. These issues are
addressed in \cite{rap04bk}, where the software required for many different
types of MD simulations is described in detail. Additionally, the case
studies described here use optimized MD software based on algorithms
customized for GPU use \cite{rap11} that, in some respects, differ
significantly from their conventional counterparts. These GPU techniques
have been extended to handle the specialized needs of the current work,
including rigid bodies, various kinds of complex boundary conditions,
multiple particle species, various force types and external fields, and have
also been updated to utilize new hardware capabilities in the latest
generation of GPUs not available for the earlier work (the actual GPU used
here is the NVIDIA K20, whose massively parallel set of 2496 computational
cores is fully utilized in the simulations).

Typically, relatively large systems and long runs are required to cover the
multiple length and time scales intrinsic to each of the systems. A run must
follow the entire evolution of the simulation, from its initial state until
the expected (or unexpected) behavior has had the opportunity to develop and
stabilize. Because local `knowledge' must be allowed to propagate throughout
the system, often a slow process that proceeds at diffusive rates, the
length and time scales are sometimes coupled, so bigger systems may need to
be simulated for longer time periods, further increasing the rate at which
the computational burden increases with system size. The dynamics are driven
entirely by deterministic interactions; the only stochastic element in the
simulations is the set of random initial velocities (and also the particle
species designation if multiple species are involved).

The particle trajectories themselves provide the raw data from which the
results are derived, such as the flow and temperature fields that can be
extracted by suitable coarse-grained averaging, the search for assembled
structures by means of cluster analysis performed on particle
configurations, or simple thermodynamic measurements. Depending on the
problem, snapshots of either the particle positions or the flow fields are
recorded over the course of each simulation, and subsequent quantitative
analysis is based on these recorded summaries. Interactive visualization
capabilities are incorporated into each of the simulations, allowing
progress to be monitored, and also enabling subsequent replay of the
recorded snapshot files.

Each model involves adjustable parameters, some of which are crucial to
determining behavior, others less so. Certain parameters correspond directly
to macroscopically measurable quantities, such as density, temperature or
gravitational field, with some parameter combinations yielding dimensionless
values such as the Reynolds or Rayleigh numbers, while other parameters are
more closely associated with the model itself, such as the interaction
strengths, the components of the friction forces, particle-wall interactions
and particle shape. Each such parameter contributes to the overall behavior,
either in the same manner as its macroscopic analog, if one exists, or
possibly in a less obvious manner if it only appears in the definition of
the model; consequently, even if the model is correctly designed in
principle, emergent behavior might only manifest itself over limited
parameter ranges. The relation between macroscopic quantities and model
parameters is obvious in some cases, such as temperature which is
proportional to mean kinetic energy, less obvious in others such as the
Reynolds number which is useful only when the fluid viscosity does not vary
significantly, an assumption not necessarily true under MD flow conditions,
and difficult to resolve as in the case of macroscopic rate constants for
self-assembly which, although dependent on interaction strength and
concentration, can have as many distinct values as there are assembly steps.

An important issue bearing on simulations of complex behavior is
reproducibility. This feature is no less vital in simulation than in
experiment, and is not usually a problem, especially when simply measuring
averages in a system already in a steady state. However, with simulations
whose outcomes are `singular' events (typically related to the emergent
aspect of the behavior), and where computational cost limits sample coverage
based on multiple runs with different initial conditions, it is not always
possible to satisfy this requirement. In each such run, these events (here
they would be associated with the stages of flow development, spatial
segregation or cluster formation) can lead to changed final states, or can
appear at different times, or sometimes even fail to appear at all.
Moreover, a run started from a particular initial state may itself have
limited reproducibility given that different computer hardware and software
combinations can produce minute changes to floating-point results that grow
exponentially with time, a well-known effect due to the underlying chaotic
nature of MD trajectories, leading to altered collective behavior (the same
is true in experiment, though for other reasons). Finally, with modern
parallel computers (GPUs in particular), the computation order itself may be
beyond user control (known as a race condition), leading to similarly
irreproducible outcomes. However, if the system state is not too close to a
phase transition (in the most general sense), the results can be expected to
be similar, even if some of the less-important details vary.

\section{Atomistic fluid dynamics}

MD provides, at least in principle, a direct approach to the study of fluid
dynamics and the variety of complex flows associated with hydrodynamic
instability, such as those illustrated in \cite{van82}. Due to its heavy
computational demands, MD cannot compete with well-established continuum
techniques (nor with lattice-based methods), but it does offer a means for
overcoming issues introduced when quantities commonly represented by
continuous fields need to allow for features such as free surfaces, multiple
components, and history-dependent behavior associated with complex rheology.
On the other hand, given the restriction to relatively small systems over
limited time intervals, there is the question of whether the familiar
continuum behavior is relevant at all in MD work; only the simulations
themselves are capable of providing an answer. An early step in bridging the
conceptual gap between atomistic fluids and the continuum was an MD study of
long-time effects arising from correlations in the atomic trajectories
\cite{ald70}.

Much of conventional hydrodynamics is governed by dimensionless parameter
combinations such as the Reynolds and Rayleigh numbers; in some cases MD
allows systems to achieve the values needed to produce the associated
laminar flow instabilities, but to compensate for the small size the
conditions imposed, such as a high shear rate or temperature gradient,
differ by orders of magnitude from the experimental regime. Even the
transport coefficients can vary across the system (due to density and
temperature variations), unlike experiment where they are typically regarded
as constant (as in Newtonian fluids). Care is needed to avoid artifacts due
to these extreme conditions. An additional complication for MD is the need
to counteract viscous heating and maintain nonslip boundaries, all without
adversely affecting the phenomena under study.

In view of these complicating factors, the expected outcome of an MD-based
fluid simulation ranges from nothing interesting if the system is too small
or the times scales inaccessible, transient effects if flows cannot be
sustained because of inadequate contact with the boundaries or coupling to
the driving field (resulting in an inability to establish nonslip flow or an
appropriate temperature gradient), and totally unrelated behavior if the
conditions are too extreme. Given the many possible failure modes, the
appearance of behavior that is qualitatively, and even quantitatively
correct, offers a pleasant surprise.

The initial applications of MD to questions of hydrodynamic instability
focused on phenomena for which a less computationally demanding
two-dimensional version could be studied, in particular, the vortices that
form in flow past a bluff body \cite{rap86b, rap87c} and Rayleigh--B\'enard
convection \cite{mar87,rap88a,puh89,rap92c,rap92d}. The two systems
discussed here are three-dimensional: Taylor--Couette flow that only exists
in three dimensions, and the more realistic three-dimensional version of the
Rayleigh--B\'enard problem \cite{cha61}. Other examples for which the MD
approach has proved successful include the Rayleigh--Taylor phenomenon
\cite{kad04} and droplet formation during the breakup of fluid jets
\cite{mos00}. In all applications of MD to hydrodynamics, fluid flow is
superimposed on the thermal motion of the atoms, the latter entirely absent
from the continuum representation; there is a further difference between the
two problems considered here, namely that vortex flow in the Taylor--Couette
system, itself a consequence of the instability caused by an adverse angular
momentum gradient, is superimposed on the overall azimuthal flow, while in
the Rayleigh--B\'enard system, where the instability is due to an adverse
temperature gradient, convection is the only flow present.

\subsection{Taylor--Couette vortices}

\begin{figure}
\begin{center}
\includegraphics[scale=0.45]{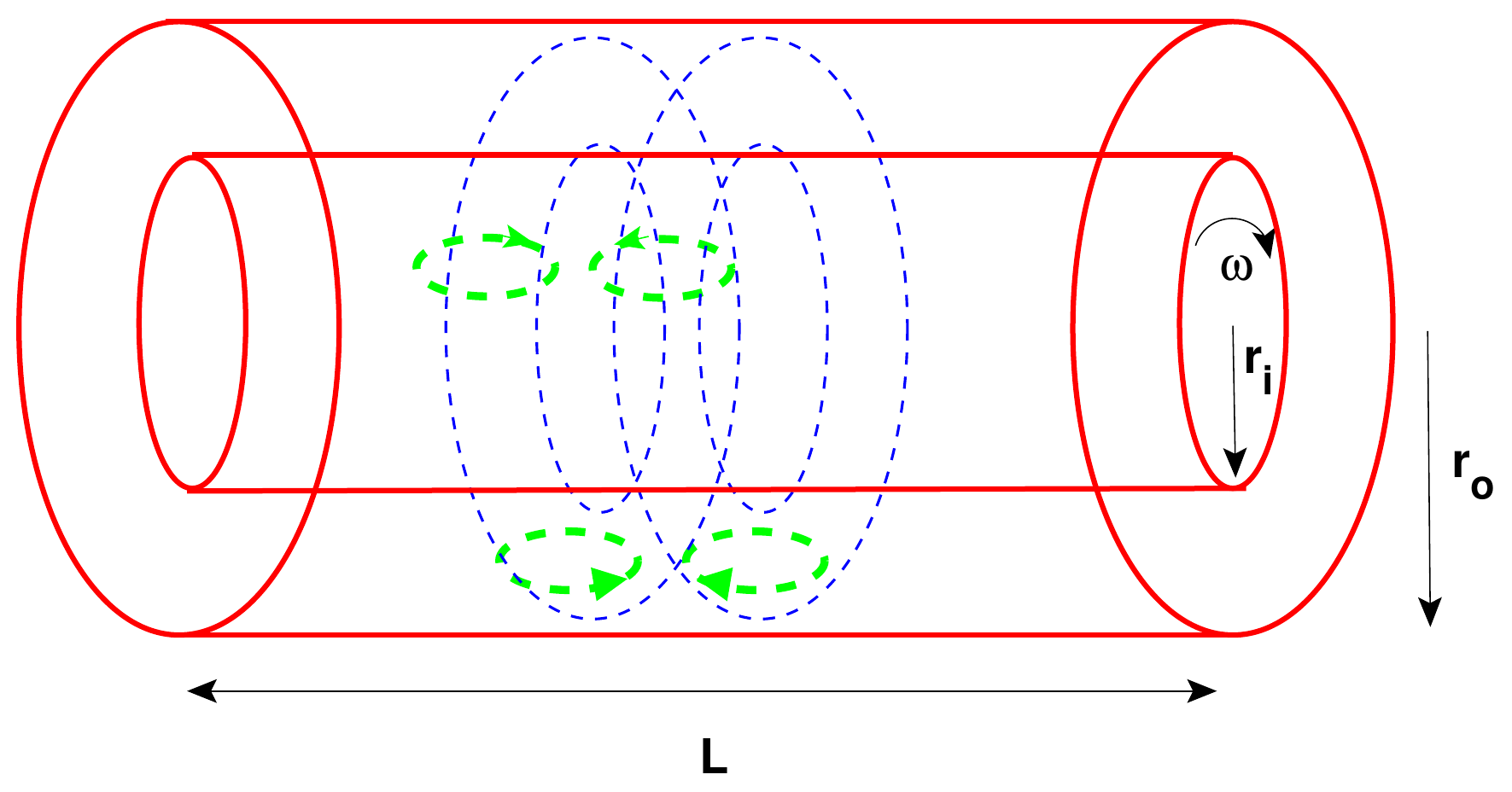}
\end{center}
\caption{\label{fig:mdtayc_schem}
Schematic diagram of the Taylor--Couette cell in which the fluid is confined
to the annular region and the inner cylinder rotates; counter-rotating
vortex rolls (such as those shown) are superimposed on the overall azimuthal
flow.}
\end{figure}

The formation of toroidal vortices in a fluid confined to an annular gap
between concentric rotating cylinders, known as the Taylor--Couette
phenomenon \cite{tay23,cha61,dra81}, is arguably the most thoroughly studied
of hydrodynamic instabilities. Consider the experimental cell defined in
Figure~\ref{fig:mdtayc_schem}, where $d = r_o - r_i$ is the width of the
annulus. For the case of a stationary outer cylinder and an inner cylinder
rotating with angular velocity $\omega$, the flow depends on the
dimensionless Taylor number
\begin{equation}
Ta = 4 [(r_o - r_i) / (r_o + r_i)] \, Re^2
\end{equation}
where
\begin{equation}
Re = d r_i \omega / \nu
\end{equation}
is the Reynolds number and $\nu$ the kinematic viscosity. The flow is purely
azimuthal at low $Ta$, but at a critical $Ta_c$, whose value can be computed
from a stability analysis of the Navier-Stokes equations \cite{cha61},
secondary flow patterns appear that have the form of regularly spaced
axisymmetric vortices. The wavelength of this pattern, corresponding to the
axial length of a pair of counter-rotating vortices, is close to $2 d$. At
higher $Ta$ \cite{col76,and86,dip81} additional azimuthal wave instabilities
appear and eventually turbulence. Numerical results using the conventional
continuum approach are described in \cite{mar84, luc85}. Experimental
studies have employed a variety of measurement techniques in examining
different aspects of the flow; examples include
\cite{don65,sny66,gol76,ber86,hei88,wer94}.

Several additions to the standard MD approach are needed to formulate an
atomistic simulation of the Taylor--Couette system. The atoms themselves are
soft spheres whose interactions are defined in (\ref{eq:ssint}). The sheared
flow is driven by the curved cylinder walls which act both as nonslip
boundaries and as thermal sinks for dissipating the heat produced by the
shear \cite{hir98}. The walls themselves are smooth; the mechanism for
achieving an effective nonslip condition is that when an atom reaches the
wall it bounces back with a new velocity whose direction is randomly chosen
to be either reflection or reversal (the choice is actually deterministic,
but there is a sensitive dependence on the impact position that yields a
seemingly random outcome), and whose magnitude is fixed to achieve the
desired temperature; at the inner wall the rotation velocity is added. The
end caps of the region reflect atoms elastically. The computational effort
is reduced by considering only a single quadrant of the annular region, with
customized periodic boundaries, in which both coordinate and velocity
vectors are suitably transformed, used to account for the remainder of the
system during the force computations.

Flow studies require coarse-grained averaging to filter out thermal
fluctuations and produce snapshots of the spatially varying flow field. The
simulation region is subdivided into a grid of cells and the average
properties, namely the velocity vector components, velocity magnitude and
cell occupancy, evaluated for each cell over a short time interval. The
degree of coarse-graining, namely the cell dimensions and the averaging
interval, is determined by the conflicting demands of resolving the fine
spatial and temporal details of varying flow patterns while suppressing the
effect of local fluctuations. The resulting cell-averaged flow magnitude and
direction, temperature and density retain no details of the actual particles
from which they are derived, and in this respect their information content
is equivalent to the continuum fluid description.

\begin{figure}
\begin{center}
\includegraphics[scale=0.4]{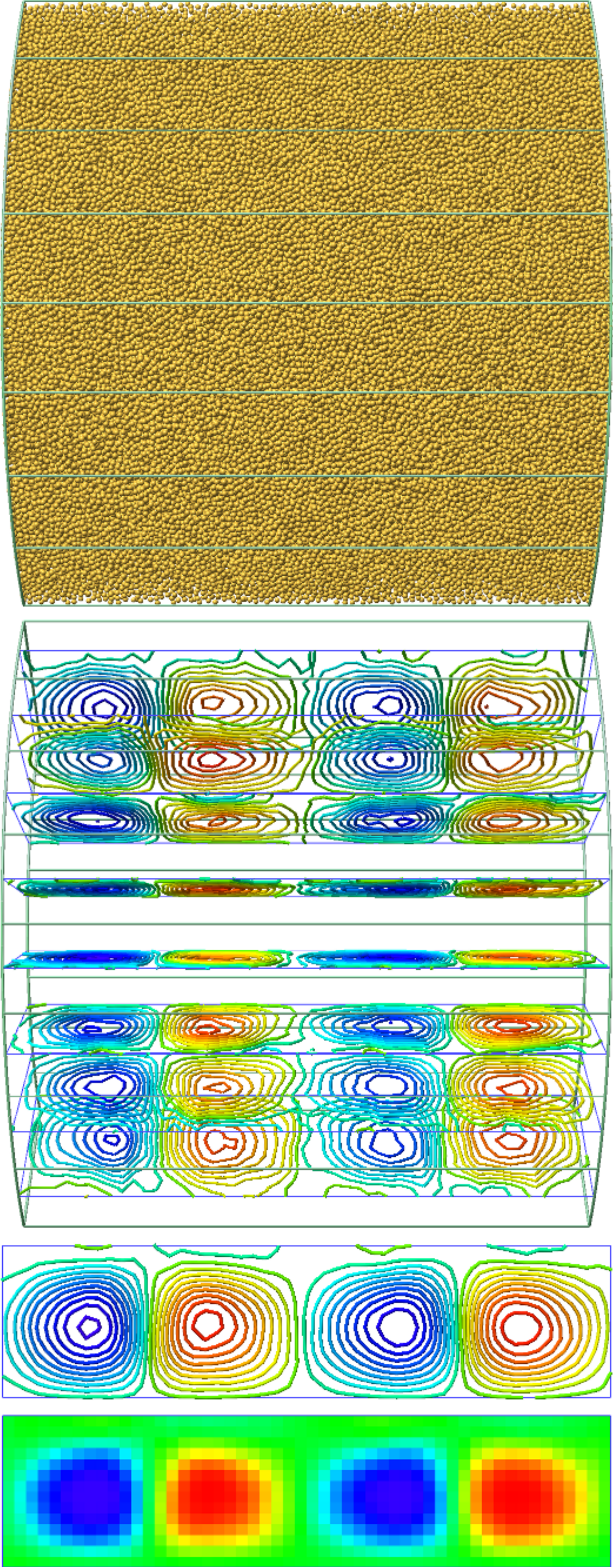}
\end{center}
\caption{\label{fig:mdtayc_small_all}
Alternative views of the smaller of the simulated Taylor--Couette systems
(after 10 revolutions): a view of the atoms filling the annular quadrant
(without the walls), streamlines based on the axial and radial flow
components in several planes at different azimuthal angles, with color
indicating the direction and speed of roll rotation, the same streamlines
averaged azimuthally, and the corresponding color plot.}
\end{figure}

Earlier work demonstrated the ability of MD simulation to not only reproduce
the Taylor--Couette vortices in a quantitatively correct fashion
\cite{hir98}, but also the rate at which these secondary flow structures
appear as a function of $\omega$ (or $Ta$) \cite{hir00}, allowing a detailed
comparison between experiment, theory and simulation.
Figure~\ref{fig:mdtayc_small_all} shows several alternative views of a
relatively small system, similar to that considered in \cite{hir98}, with
$r_i = 50$, $d = 25$, $L = 100$, containing $N \approx 1.2 \times 10^5$
atoms, for $\omega = 0.1$; $Ta \approx 1.6 Ta_c$ for this system. The first
of the views shows the actual MD system; the azimuthal component of the flow
can be seen when the simulation is run interactively, superimposed on the
thermal motion of the particles, but the fact that the averaged particle
trajectories are helical, a consequence of the toroidal rolls, can only be
inferred from the coarse-grained flow analysis. The next view shows the flow
streamlines for several axial slices, after removing the azimuthal flow
component; the vortex structure is now apparent, with color used to
distinguish between counter-rotating flows. Since the final state is
azimuthally symmetric (this is not true during early vortex development),
full azimuthal averaging for further noise reduction leads to the
streamlines in the next image, and finally the same information in a
simplified, color-coded form (used later). These alternative views also
exemplify the different kinds of information that appear as the level of
representation is changed, ranging from excessively detailed particle motion
to the spatially averaged flow of the continuum fluid, additionally
simplified by symmetry.

The value of $d$ used in \cite{hir98} is practically the smallest permitting
vortex development. Larger systems are clearly desirable to reduce
finite-size effects; increasing the annular width $d$ provides more room for
roll growth, and if $L / d$ is also enlarged additional rolls can be
accommodated. Recall, however, that even the biggest systems addressable by
MD remain extremely microscopic. The system described here has $r_i = 100$,
$d = 40$, $L = 640$, providing a 20-fold size increase to $N \approx 2.35
\times 10^6$. The extreme conditions to which the fluid is subjected, the
shear rate being the most prominent, are alleviated slightly by reducing
$\omega$ to 0.04. Overall, these changes increase $Ta$ by a factor of 1.4
and provide a fourfold increase in $L / d$.

\begin{figure}
\begin{center}
\includegraphics[scale=0.23]{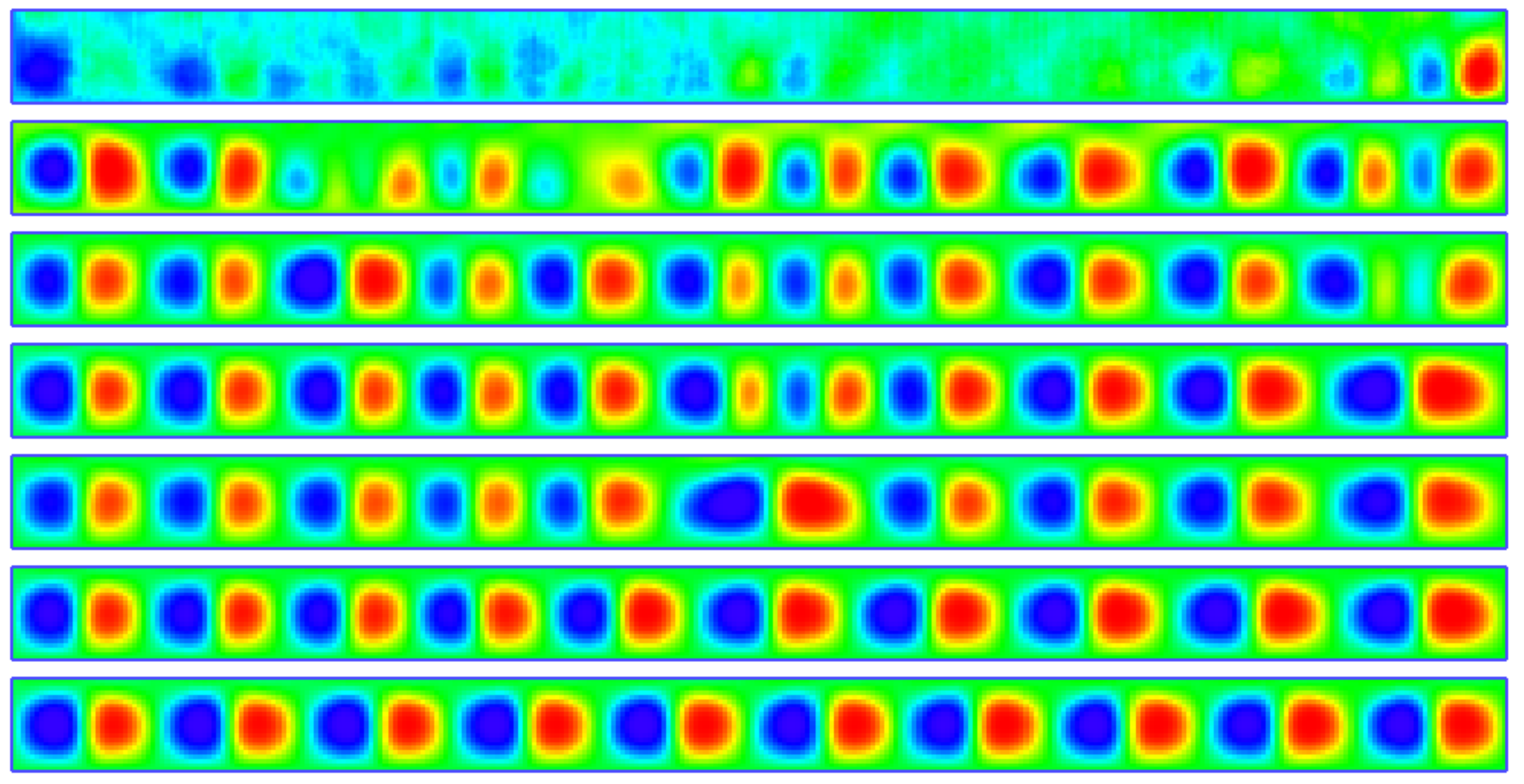}
\end{center}
\caption{\label{fig:mdtayc_stream_evol}
Azimuthally averaged color plots for the larger Taylor--Couette system at
different times (after 1.4, 2.7, 4, 9, 10, 40 and 650 revolutions) showing
the onset of the rolls and their subsequent merging and repositioning.}
\end{figure}

The toroidal vortex evolution in this larger system over a run of $1.9
\times 10^7$ timesteps, amounting to 650 revolutions, is shown in
Figure~\ref{fig:mdtayc_stream_evol}. Rapid initial vortex formation appears
within just four revolutions, covered by the first three frames of the
sequence, followed by vortex merging and the disappearance of an adjacent
pair of vortices after about ten revolutions in the next two frames. The
last two frames show gradual vortex resizing leading to a final state that
exhibits apparent long-term stability. Pattern development is not
necessarily reproducible between runs, and where the final state consists of
a sufficiently high number of vortices, even the existence of a preferred
wavelength cannot fully constrain the number of vortices present in the
final state; indeed, here there are 20 vortices, instead of 16 expected from
the theoretical pattern wavelength. Extra vortices, both transient and
permanent, are sometimes observed in the smaller system as well
\cite{hir00}; analogous effects are also encountered experimentally
\cite{ben82}.

\begin{figure}
\begin{center}
\includegraphics[scale=0.80]{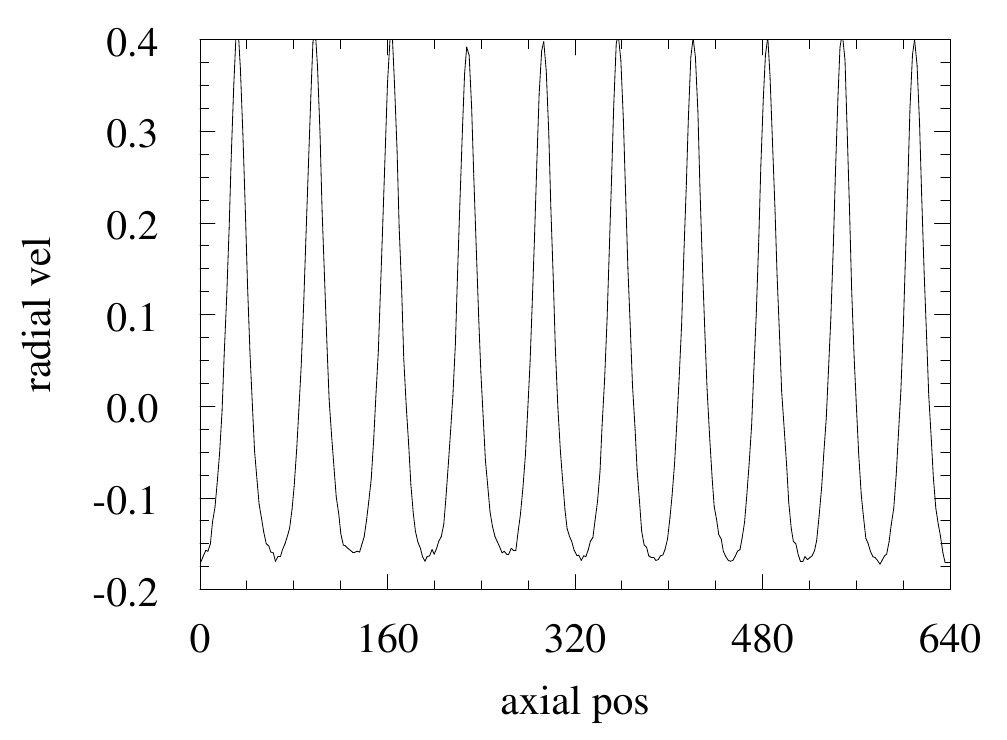}
\end{center}
\caption{\label{fig:mdtayc_rvel_plot}
Axial dependence of the radial component of the flow velocity at the end of
the Taylor--Couette simulation; each peak corresponds to a vortex pair.}
\end{figure}

The radial component of the flow (azimuthally averaged) at the end of the
run of the larger system is shown in Figure~\ref{fig:mdtayc_rvel_plot}. Each
of the 10 peaks along the cylinder axis corresponds to a vortex pair, with
the outward flow faster than the inward; for the small system there are just
two peaks, and \cite{hir98} shows that even though the behavior is strongly
nonsinusoidal, it can be fit to a three-term Fourier expansion with
coefficients predicted by theory \cite{dip81,dav62}. The maximum radial
velocity $\approx 0.4$, much smaller than the innermost value of the
coarse-grained azimuthal flow, measured to be 3.8 ($\approx r_i \omega =
4$); the 10-fold difference explains why rolls cannot be observed directly
without averaging.

\subsection{Rayleigh--B\'enard convection cells}

\begin{figure}
\begin{center}
\includegraphics[scale=0.40]{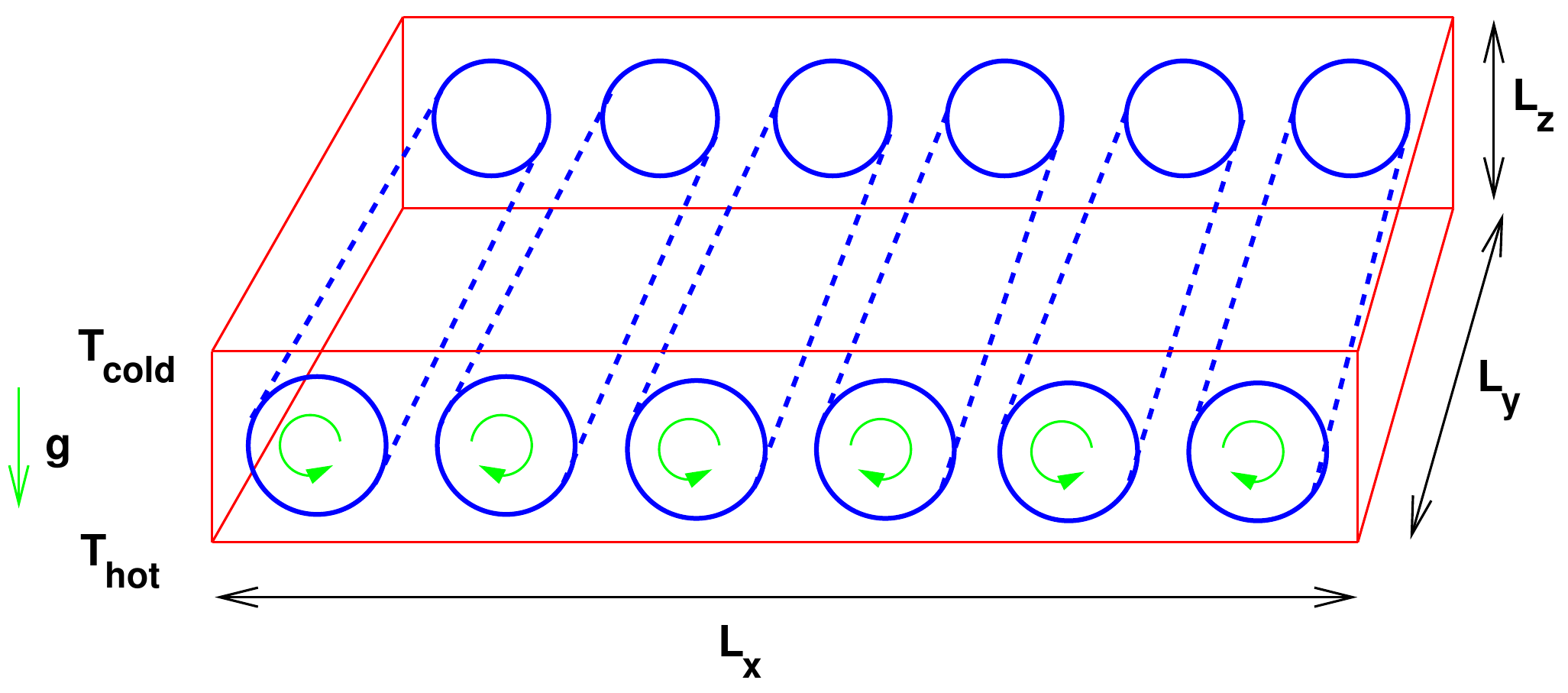}
\end{center}
\caption{\label{fig:mdrayb_schem}
Schematic diagram of the Rayleigh--B\'enard system; linear convection rolls
are shown for illustrative purposes.}
\end{figure}

The Rayleigh--B\'enard phenomenon -- the rich variety of flow patterns
produced by convection in a fluid layer heated from below -- is another
well-known example of hydrodynamic instability
\cite{cha61,nor77,dra81,ber84,kos93,bod00}. The system is defined in
Figure~\ref{fig:mdrayb_schem}, and its behavior is governed principally (but
not entirely) by the dimensionless Rayleigh number
\begin{equation}
Ra = \alpha g L_z^3 \Delta T / \nu \kappa
\end{equation}
where $\alpha$ is the thermal expansion coefficient, $g$ the gravitational
acceleration, $\Delta T = T_{hot} - T_{cold}$ the temperature difference,
$\nu$ the kinematic viscosity, and $\kappa$ the thermal diffusivity (the
other relevant ratio, not discussed here, is the Prandtl number $Pr = \nu /
\kappa$). A critical value $Ra_c$ marks the onset of convection, where
buoyancy overcomes viscous drag, and convection replaces conduction as the
principal mechanism for thermal transport. Theory is simplified by the
Boussinesq approximation, in which it is assumed that density is the only
temperature-dependent fluid property; $Ra_c$ can then be computed for
various kinds of boundary conditions \cite{cha61}, as can the wavelength of
the convection pattern at criticality, but not the predicted planform, such
as linear or concentric rolls, spirals, and hexagonal cells. The actual
convection roll (or cell) width, typically of order $L_z$, represents a
compromise: narrow rolls reduce shear losses at the nonslip walls, while
wide rolls reduce viscous drag and diffusive heat transfer between
oppositely moving hot and cold streams. The continuum version of the problem
has been studied computationally, e.g., \cite{gol89,gel99,wat97}.

\begin{figure}
\begin{center}
\includegraphics[scale=0.44]{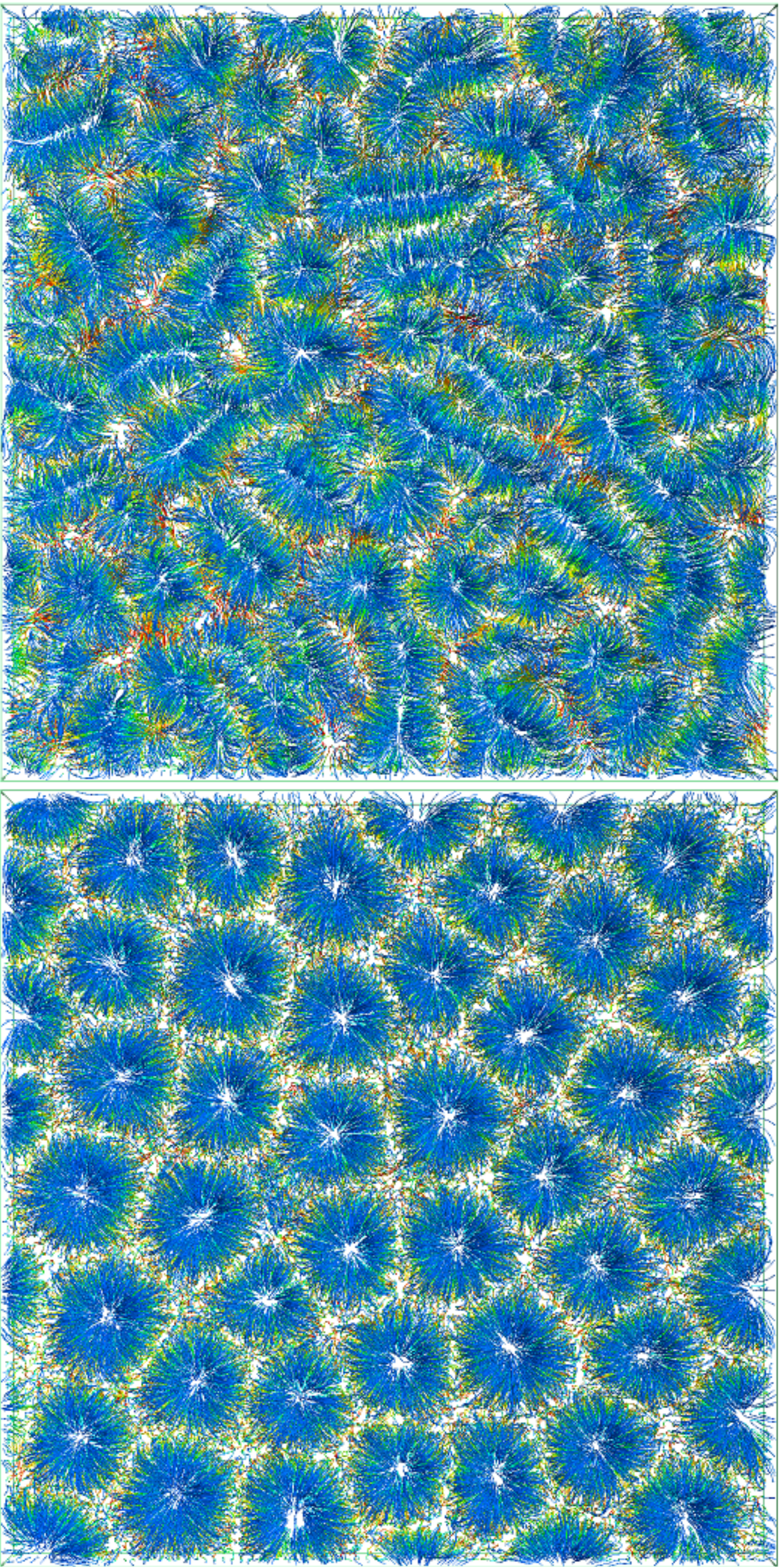}
\end{center}
\caption{\label{fig:mdrayb_stream_both}
Streamlines from the larger Rayleigh--B\'enard simulation early in the run
where organized flow structures start to develop, and at the end of the run
where there is a well-developed convection cell array.}
\end{figure}

In order to simulate the Rayleigh--B\'enard system at the atomistic level
the standard MD approach must be augmented by a mechanism for heat injection
at the bottom wall and removal at the top, as well as a gravitational field.
Coarse-grained flow analysis is once again required to examine the
developing flow. The simulations described in \cite{rap06} deal with a
system of $N \approx 3 \times 10^6$ atoms, together with an additional
$\approx 0.5 \times 10^6$ fixed atoms that form the bottom and top walls. A
run of $3.2 \times 10^6$ timesteps (of size 0.004), with $T_{hot} = 10$,
$T_{cold} = 1$, and $g$ arbitrarily set to $3 \Delta T / 2 L_z$ (a choice
ensuring equality of the potential and kinetic energy changes across the
layer), resulted in a periodic array of eight hexagonal convection cells. A
narrower system with smaller $L_y$ produced linear rolls of the type shown
schematically in Figure~\ref{fig:mdrayb_schem}.

The system considered here has three times the number of atoms, with region
size $L_x = L_y = 679$ and $L_z = 54.3$ (compared with the earlier $L_x =
521$, $L_y = 451$ and $L_z = 35.3$ \cite{rap06}), and $N = 10^7$. $T_{hot}$
is lowered to 8 (for a less `extreme' thermal gradient) and $g$ is halved to
reduce density variation across the layer. These changes increase the
nominal value of $Ra$ by 10\%; since $Ra$ depends on $\nu$ and $\kappa$,
both of which vary with local temperature and density, the actual value of
$Ra$ is not readily determined (even its relevance under the extreme
non-Boussinesq conditions is questionable). While the lateral boundaries are
periodic as before, the bottom and top thermal walls are implemented using a
new mechanism, avoiding the need for additional fixed atoms: Fluid atoms
colliding with these horizontal, planar walls are assigned new velocities,
with a magnitude set by the nominal wall temperature and direction
corresponding to a random choice between reflection or reversal (the choice
actually depends on the impact position, as in the Taylor--Couette
simulations); in addition (following \cite{rap06}), the velocities of all
atoms close to the thermal walls (within $\approx r_c$) are rescaled at
regular intervals, after zeroing their transverse average, to avoid any
buildup of unwanted drift and help enforce the desired boundary
temperatures. This change, as will become apparent from the results, leads
to the development of narrower convection cells than before, in closer
agreement with the expected behavior.

The outcome of the simulation after $10^7$ timesteps, requiring an order of
magnitude more computation than previously, is a much more elaborate
convection cell pattern. Streamlines are used in
Figure~\ref{fig:mdrayb_stream_both} to show the nascent flow state after
just $4 \times 10^4$ timesteps, and the reasonably well-ordered cell array
at its end. The color-coded streamlines show temperature variation, and
while the downward flow of cold fluid at the cell centers is clearly visible
when viewed from above, the upward flow of hot fluid at the cell edges is
best seen when the system is viewed obliquely or from below (not shown). The
measured vertical component of the flow velocity is in the approximate range
(-1.8, 2.6), where the nominal thermal velocities are $\approx 1.7$ and 5 at
the cold and hot boundaries respectively; at mid-height the velocity range
is reduced to (-1.0, 2.0).

\begin{figure*}
\begin{center}
\includegraphics[scale=0.30]{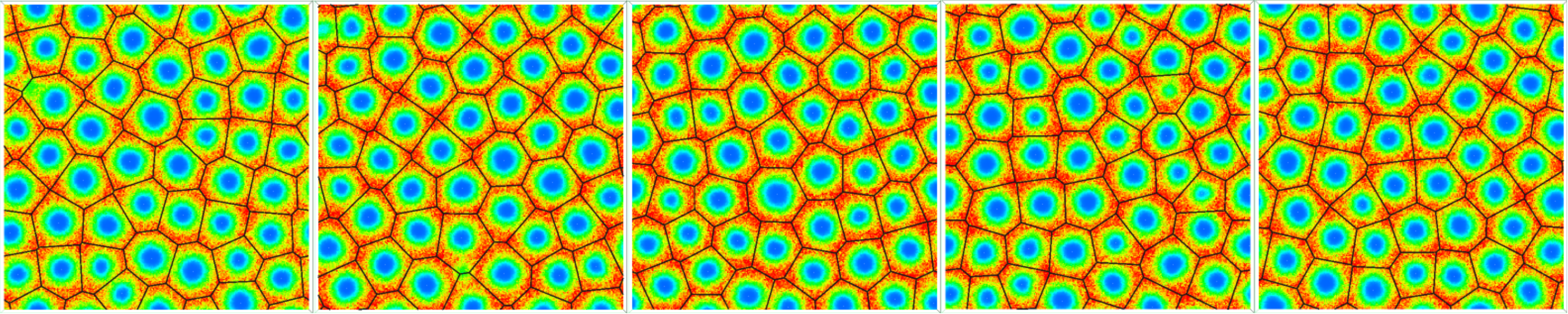}
\end{center}
\caption{\label{fig:mdrayb_vor_all}
Variation of the convection cell pattern with time (larger
Rayleigh--B\'enard system, at times 0.7, 1.7, 2.7, 3.7 and $5 \times 10^4$),
with automated Voronoi analysis used to delineate cell boundaries.}
\end{figure*}

The early convection pattern contains certain linear features, but
conversion into compact cells is complete after $\approx 3 \times 10^5$
timesteps, following which there is a slow process of cell resizing and
rearrangement; a series of snapshots showing how the cell pattern changes
with time appears in Figure~\ref{fig:mdrayb_vor_all}. Cell organization
changes gradually, with no clear indication that the process eventually
terminates; while the number of convection cells is almost constant (see
Figure~\ref{fig:mdrayb_cells_plt} below), new cells are seen to appear and
existing cells merge, with each such event followed by rearrangement of
nearby cells. It is worth bearing in mind that each convection cell
represents coherent flow involving $\approx 2 \times 10^5$ MD fluid atoms
superimposed on the random thermal motion, an impressive exhibition of
emergent collective behavior.

Quantitative cell analysis is based on coarse-grained temperature
measurements at a height $0.3 L_z$ above the hot bottom wall, where the
variation is most pronounced. The cell centers are positioned at each of the
local temperature minima (coincident with the downward flow maxima) and
Voronoi polygon analysis (a simplified two-dimensional version of the
technique described in \cite{rap04bk}, in which the plane is subdivided into
polygons where the region nearer to a particular center than to any other is
assigned to its polygon) is used to predict the cell boundaries. For a
well-developed cell array these computed boundaries closely track the actual
flow cell structure.

A variety of measurements can be made based on the Voronoi polygon analysis.
The convection cell count is shown as a function of time in
Figure~\ref{fig:mdrayb_cells_plt}; the number of cells varies, but only over
a narrow range. The average number of edges per polygon must of course be
six, with an observed $\pm 1$ variation, the mean radius (specifically the
mean inradius or distance from the polygon center to its edges) depends on
the number of polygons and varies between 54 and 57, with a spread (a
measure of the deviation from a regular polygon in which all distances are
equal) that ranges from 2.3 to 12 for individual snapshots. The average
value (and standard deviation) of this radial range is included in
Figure~\ref{fig:mdrayb_cells_plt}, and the fact that it amounts to only
about 10\% of the mean radius is consistent with the observed regularity of
the cell shapes. The underlying wavelength of the cell array, $\lambda$, is
twice the mean radius, so that here the measured value of $\lambda / 2$ is
very close to $L_z$ (= 54.3).

\begin{figure}
\begin{center}
\includegraphics[scale=0.82]{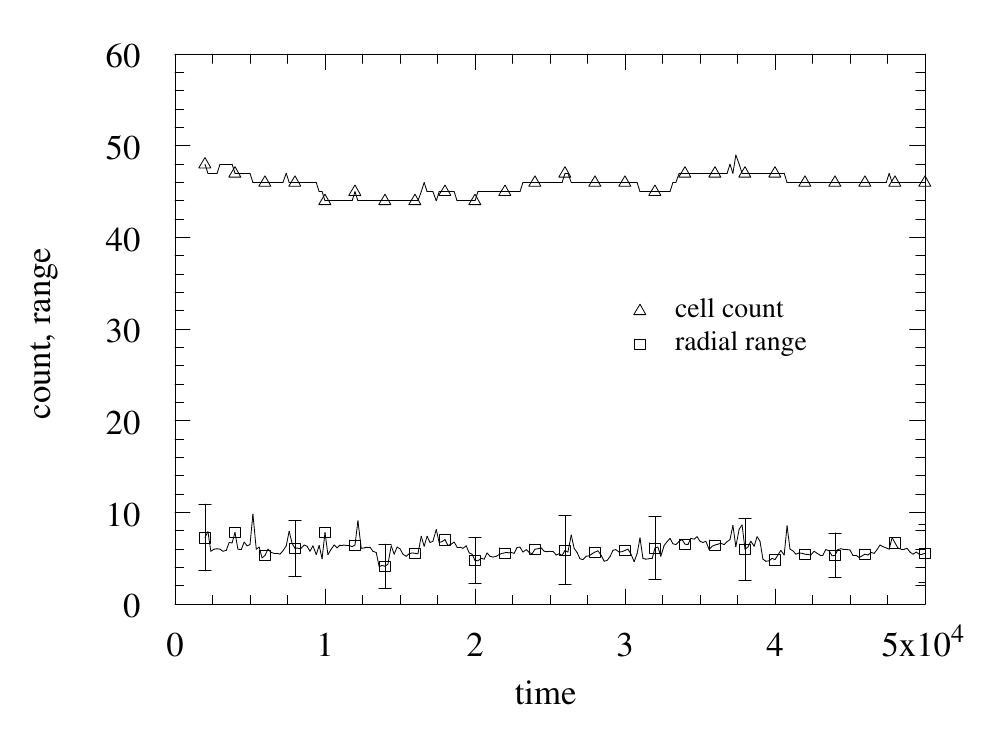}
\end{center}
\caption{\label{fig:mdrayb_cells_plt}
Number of convection cells and their average radial range over the entire
run (larger system).}
\end{figure}

\begin{figure}
\begin{center}
\includegraphics[scale=0.44]{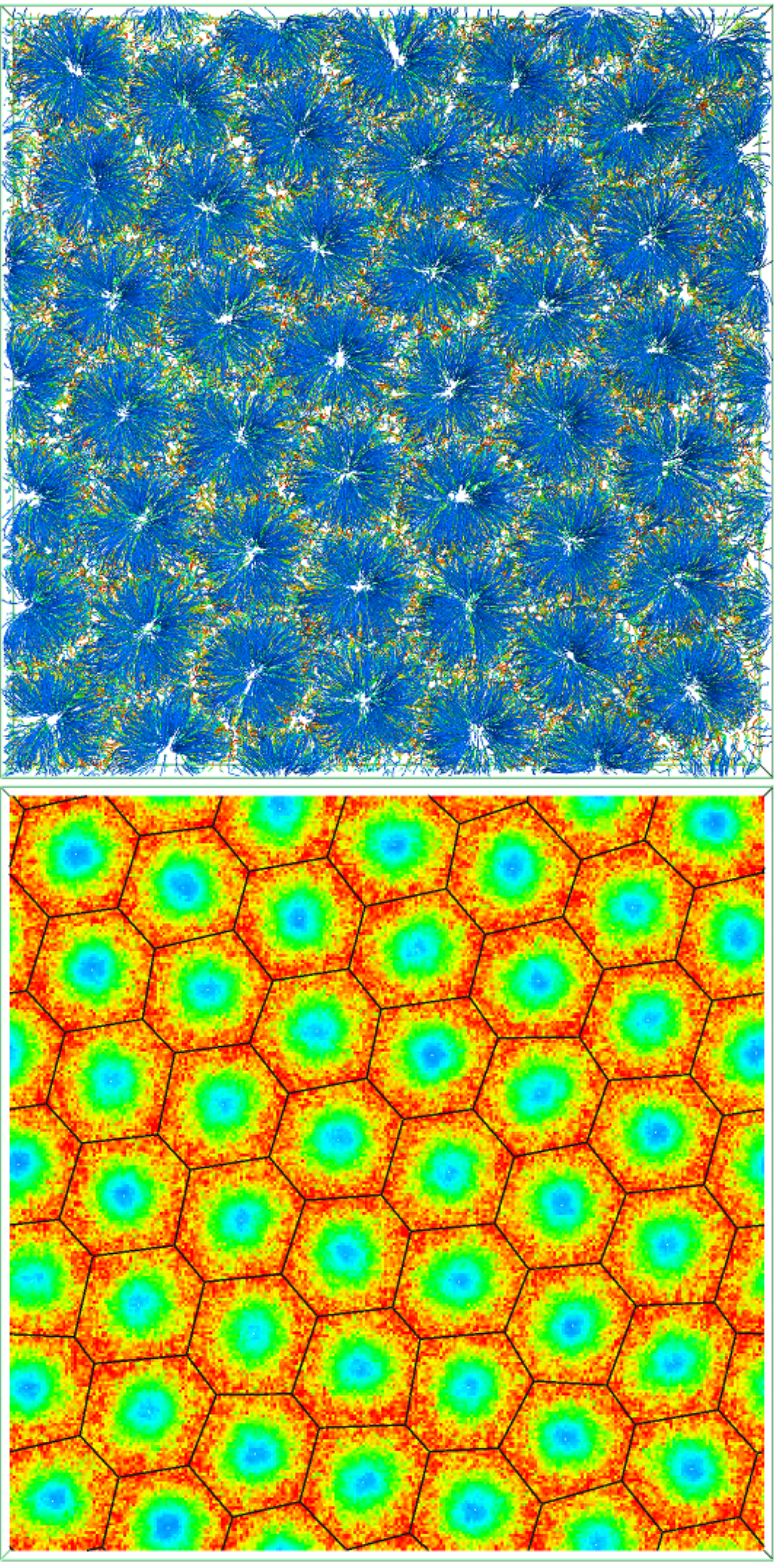}
\end{center}
\caption{\label{fig:mdrayb_med_both}
Flow streamlines and Voronoi analysis for the smaller Rayleigh--B\'enard
system showing the defect-free lattice of hexagonal convection cells.}
\end{figure}

There are considerably more convection cells in this simulation than in
\cite{rap06}, a far larger number than can be attributed to the altered
system size; in the earlier simulations $\lambda = 5.27 L_z$, but here this
has dropped to $\lambda = 2 L_z$. Theory \cite{cha61} predicts different
$\lambda$ values for hexagonal cells and linear rolls, with $\lambda_{hex} /
\lambda_{lin} = 2.14$ and $\lambda_{lin} = 2.016 L_z$; different ratios for
(gently curved) rolls and hexagons have been reported experimentally
\cite{ass96,baj97,roy02}, even down to unity. Thus the fact that the MD
hexagons have a value of $\lambda$ predicted for linear rolls is not
unreasonable since experiment implies that the simplified theory is
inadequate. The likely cause of the enlarged $\lambda$ in the earlier
simulations is the wall implementation, where relying on collisions with a
boundary array of spheres to produce the nonslip condition is less effective
than individually adjusting the post-collision velocities; the present
smaller $\lambda$, a value consistent with experiment, appears to be a
consequence of the more effective method.

The dependence of $\lambda$ on $L_z$ is confirmed by considering a smaller
but otherwise similar system. The size is reduced to $L_x = L_y = 543$, $L_z
= 40.7$ (this $L_z$ is only slightly bigger than in \cite{rap06}) and $N =
4.8 \times 10^6$; the edge ratios $L_x / L_y$ and $L_x / L_z$ are changed
and the nominal $Ra$ is reduced by a factor of 0.4. The outcome of this run
after $1.3 \times 10^7$ timesteps, seen in Figure~\ref{fig:mdrayb_med_both},
is a perfect array of hexagonal convection cells; there is no visible change
in this array over the last $3 \times 10^6$ timesteps. From a Voronoi
analysis of this run the mean polygon radius is 42.8, again very close to
$L_z$, and the mean radial range of the cells is just 2\% of the radius,
confirming the observed regularity of the hexagon array. The initial
orientation of the cells is decided spontaneously in the absence of physical
lateral boundaries, although if the final state is a perfect array the
periodic boundaries will constrain its orientation; multiple runs (with
different initial conditions) show similar overall behavior but the detailed
cell development differs. These represent initial results for this problem;
the extended cell arrays ought to be amenable to Fourier analysis, not
attempted for the very limited cell array in \cite{rap06}. Furthermore, in
view of the dependence on region size, shape and other parameters, as well
as the sensitivity to the initial state originally seen in two dimensions
\cite{rap92c}, more extensive simulations, especially near the convection
threshold \cite{rap92d}, are likely to prove interesting.

\section{Granular segregation}

Granular dynamics is a field fraught with surprise, where the mechanisms
responsible for much of the counterintuitive behavior remain a longstanding
theoretical challenge; potential economic benefits provide ample motivation
for achieving a systematic understanding of the phenomena involved. Since
the behavior often differs strongly from systems explainable by fluid
dynamics and statistical mechanics, established theory is limited in the
help it can provide and, consequently, progress
\cite{bar94,her95,jae96,kad99} relies heavily on computer simulation.

Granular segregation is a fascinating example of emergent behavior: the
ability of dry, noncohesive particle mixtures to separate into individual
species, despite the absence of any obvious energetic or entropic advantage.
Segregation and mixing are central to many kinds of industrial processing,
so that the capability of either causing or preventing species separation is
important.

Computational methods analogous to MD, with particles now corresponding to
macroscopic grains, provide a flexible approach to simulating granular
matter \cite{cun79,wal83,haf86}. Additional forces are introduced to
represent the effects of inelasticity and friction, the simplest of which is
velocity-dependent collision damping; although the approximations tend to be
empirical, since a full prescription for reproducing specific granular
properties is unknown, suitable interaction functions allow simple spherical
particles to mimic rough, irregular grains (in a vacuum). More complex
particle shapes can also be employed, as is the case in one of the examples
discussed here. Finally, since container boundaries are usually involved,
often providing the driving force, the simulations must take them into
account.

Granular simulations can have several possible outcomes, as was the case
with atomistic fluid dynamics. These range from nothing interesting if the
model fails to incorporate the essential properties of the granular medium,
incorrect (but perhaps still interesting) behavior if the model somehow
misrepresents reality, partially correct behavior in which some features are
captured, and preferably, a correct reproduction of the form and sequence of
the segregation process. With this in mind, two examples are described. One
is a granular mixture in a horizontal rotating cylinder, a system known to
exhibit various combinations of axial and radial size segregation; here,
simulation successfully imitates experiment. The other is a mixed granular
layer on a vertically vibrating, sawtooth-shaped base, where a combination of
stratified horizontal counterflows and vertical (Brazil nut) size
segregation might be expected to produce horizontal segregation; while the
simulations support this prediction, experimental verification is still
awaited.

\subsection{Segregation in a rotating cylinder}

Segregation of a binary mixture in a partially filled cylinder rotating
around a horizontal axis is an extensively studied phenomenon \cite{sei11}.
Axial segregation is the most prominent effect, where a pattern of bands of
alternating particle species develops along the axial direction; bands can
gradually merge, at a rate that drops to such a low level that it is unknown
whether the state eventually reached is stable, or just longlived. A second
form of segregation, requiring greater effort to observe, occurs in the
radial direction, resulting in a core rich in small particles surrounded by
a layer of predominantly big particles; the radial core itself can be the
final state, a transient state preceding axial segregation, or a feature
which persists even when axial segregation is apparent externally.

Early experimental efforts involved direct observation of axial band
structure from the outside \cite{zik94}; some of the subsequent studies
succeeded in examining the interior using MRI \cite{hil97f,nak97} and
relating the development of the axial bands to bulges occurring in the
radial core. Time-dependent behavior can occur after the initial appearance
of the axial bands, with gradual merging of narrower bands \cite{fre97} and
traveling surface waves \cite{cho98}. Complicating the behavior are results
showing that the ratio of cylinder to particle diameter determines if axial
segregation is possible and whether its appearance depends reversibly on
rotation rate \cite{ale04}. Later experiments \cite{arn05,fin06} revealed
more about the richness of the segregation effect and the associated
dynamics under wet conditions; these are relevant because slurries tend to
share the behavior of dry mixtures. Overall, no consensus has yet emerged as
to the mechanisms underlying this form of segregation; the difficulties
inherent in understanding granular matter are discussed in
\cite{jae96,kad99,ara06}.

Granular simulations have proved capable of reproducing both the axial and
radial forms of segregation. In \cite{rap07a} it was shown how the
appearance of radial segregation precedes axial segregation, and that
depending on the circumstances, a radial core of small particles could
persist even after the external view showed essentially complete axial
segregation. Corresponding behavior in three-component mixtures was
described in \cite{rap07b}. Owing to the need for relatively large systems
and long runs, only a very limited set of parameter combinations could be
considered. The problem of following the development of segregation in
detail is further exacerbated by a lack of reproducibility between runs that
differ only in their initial states (i.e., the random initial velocities
assigned to the particles) leading to varied segregation scenarios; the same
is true experimentally, implying that multiple simulation runs are needed
for observing typical behavior, a situation similar to the fluid studies
discussed earlier.

\begin{figure}
\begin{center}
\includegraphics[scale=0.25]{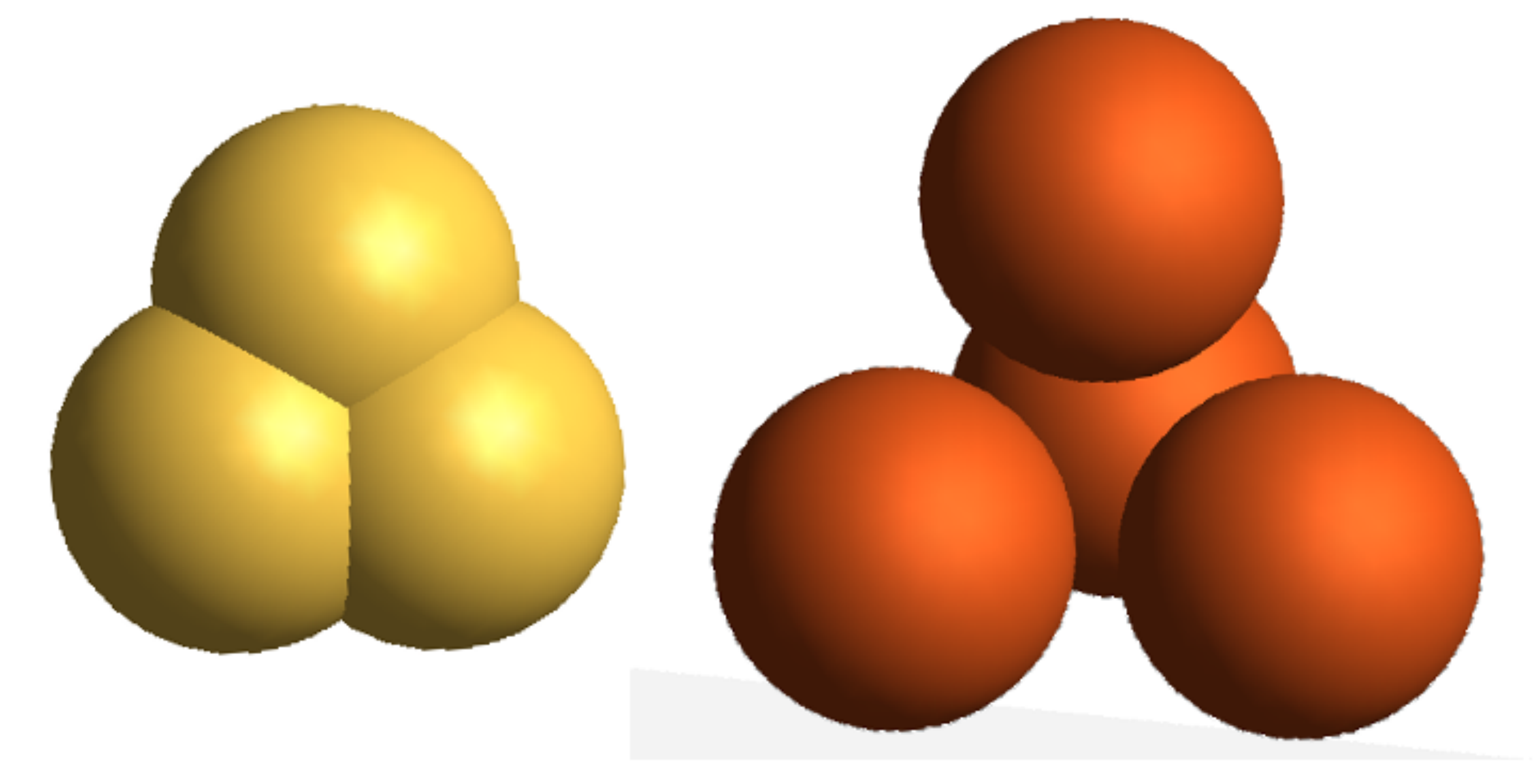}
\end{center}
\caption{\label{fig:grcyl_particles}
Particles used in the cylinder granular segregation simulations; each
consists of four spheres in a rigid tetrahedral array (the virtual
`framework' is not shown), the only difference between big and small
particles is the distance between spheres.}
\end{figure}

Increased computer performance allows this problem to be revisited, but with
several changes to the model. In \cite{rap07a} soft-sphere particles were
used, together with the usual velocity-dependent normal and tangential
forces for representing collision damping and sliding friction, defined in
(\ref{eq:normdamp}) and (\ref{eq:tandamp}) below, as well as tangential
restoring forces (involving the position histories of the relevant particle
pairs) introduced to mimic the effect of static friction. In the new
simulations, instead of using simple spherical particles, each particle is
constructed from a set of spheres rigidly positioned on a (virtual)
tetrahedral frame, as shown in Figure~\ref{fig:grcyl_particles}; particles
of different size differ only in their tetrahedral arm lengths, in this case
0.6 and 1.2 (MD units). Unlike purely convex particles, the more complex
effective shape allows particles to interlock, thereby partially reproducing
the effect of static friction without the tangential restoring force used
previously (whose GPU implementation is inconvenient); the geometrical
interlocking effect is a simplified version of what really happens, where
interlocking (and breakage) of multiple asperities on rough grain surfaces
is the ultimate cause of static friction. The curved cylinder boundary is
also required to emulate the effects of static friction, and this is
achieved by embedding spherical particles in the surface of the rotating
wall, arranged on a staggered grid with spacing 2.2, producing a
sufficiently rough boundary to impede sliding. The cylinder end caps provide
normal damping, but are otherwise smooth. The resulting segregation effects
turn out to be unaffected by the alterations to the model.

The particle interactions are based on \cite{rap07a}, with the exception of
the omitted tangential restoring force. The repulsive force between spheres
$\avec{f}_\ell$ is obtained from the potential in (\ref{eq:ovint}), with
$k_n = 1000$. The velocity-dependent normal damping is
\begin{equation}
\avec{f}_d = - \gamma_n (\avecu{r} \cdot \avec{v}) \avecu{r} \label{eq:normdamp}
\end{equation}
where $\avec{v}$ is the relative sphere velocity and $\gamma_n = 5$. Sliding
friction is due to tangential damping between particles within contact range
\begin{equation}
\avec{f}_s = - \min ( \gamma_s |\avec{v}_t|, \mu |\avec{f}_\ell + \avec{f}_d| )
\avecu{v}_t \label{eq:tandamp}
\end{equation}
where $\avec{v}_t$ is the transverse component of $\avec{v}$; the damping
coefficients are $\gamma_s = 10$ for spheres associated with the big
particles and for sphere-wall contacts, and $\gamma_s = 2$ otherwise
(following \cite{rap07a}, $\gamma_s$ is larger for big particles); the
static friction coefficient $\mu \propto \gamma_s$, with a maximum of 0.5.
Gravity is of course present.

The two examples considered here have different cylinder dimensions and
demonstrate dissimilar segregation scenarios. One involves a shorter
cylinder of length $L = 200$ and diameter $D = 100$, the other a longer
cylinder with $L = 500$ and $D = 80$; the dimensions need to be larger than
in \cite{rap07a} because of the increased effective particle size. For the
initial state, particles are placed on a lattice filling the cylinder,
assigned small random velocities, and the species (big or small) is set
randomly; the lattice spacing is chosen to produce a specific fill level
when particles settle on the bottom (the different fill levels can be seen
in the pictures below) and in the present simulations the shorter cylinder
is filled to a higher level. The fraction of big particles is 0.33; this
ensures that when axial bands form, the total band widths for the two
species are roughly equal. The number of particles in the two systems are $N
= 1.36 \times 10^5$ and $1.03 \times 10^5$; the actual number of soft
spheres involved in the force computations (apart from the wall spheres) is
four times larger. The cylinder rotates with angular velocity $\omega =
0.2$; an upper bound to $\omega$ is set by the ratio of angular to
gravitational acceleration, the Froude number $Fr = D \omega^2 / 2 g$, to
prevent particles forming a ringlike layer adjacent to the curved cylinder
wall; here $g = 5$, so $Fr = 0.4$ and 0.32 for the two systems. If the
length unit (the sphere diameter) is $10^{-3}$\,m, then the time unit for
this value of $g$ corresponds to 0.022\,s, resulting in an experimentally
reasonable 1.4\,Hz rotation rate.

\begin{figure}
\begin{center}
\includegraphics[scale=0.30]{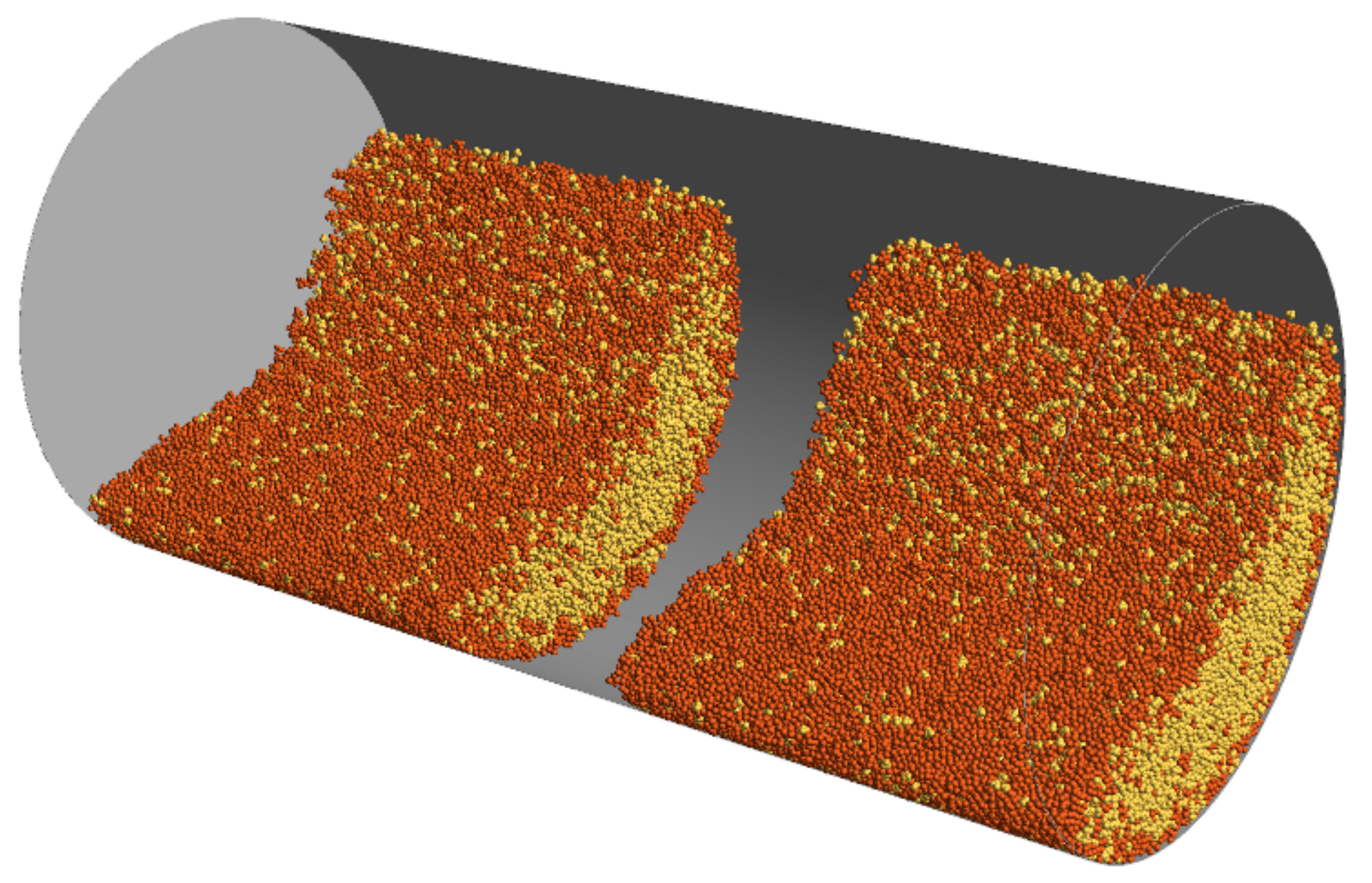}
\end{center}
\caption{\label{fig:grcyl_rad_seg}
Radial granular segregation (shorter cylinder) with big and small particles
colored red and yellow respectively; a slice has been removed to show the
interior, and particles embedded in the curved cylinder wall (to produce
surface roughness) are omitted.}
\end{figure}

\begin{figure}
\begin{center}
\includegraphics[scale=0.24]{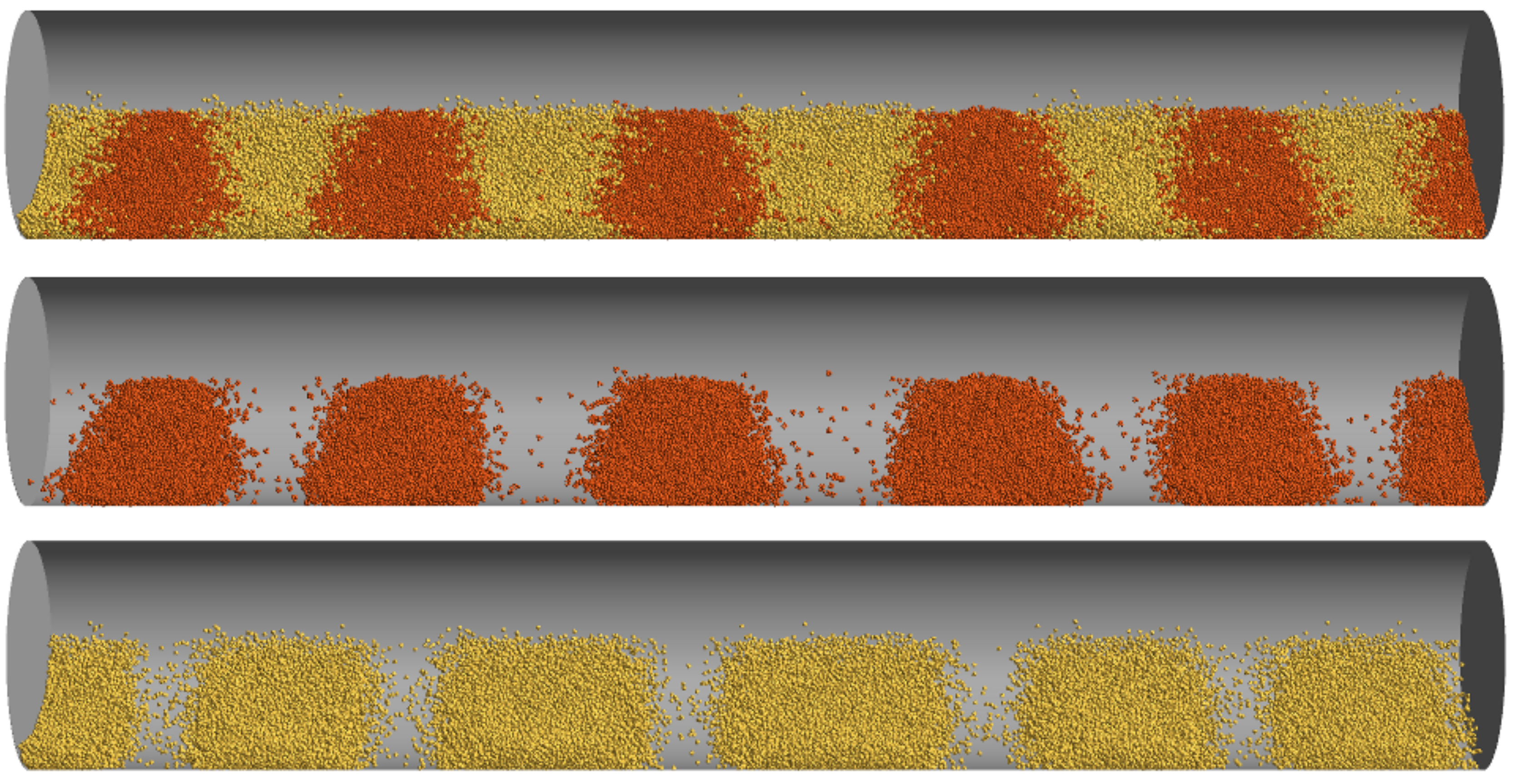}
\end{center}
\caption{\label{fig:grcyl_ax_seg}
Axial granular segregation (longer cylinder); the complete system is shown,
together with separate views of the big and small particles.}
\end{figure}

Simulation of the system with the shorter cylinder provides a demonstration
of radial segregation, as shown in Figure~\ref{fig:grcyl_rad_seg}. In this
picture, a slice has been removed so that the interior organization (away
from the end walls) can be seen; the inner core is dominated by small
particles, and the majority of big particles reside in the outer layer. The
onset of segregation is rapid, reaching completion after approximately 30
revolutions; radial segregation is also a fast process experimentally. The
simulation run extends over a total of 1120 revolutions ($7.2 \times 10^6$
timesteps) but no subsequent organizational changes are observed, a strong
indication that this state is stable. The same system run at a reduced
$\omega = 0.1$ (not shown) produces even stronger radial segregation.

\begin{figure}
\begin{center}
\includegraphics[scale=0.65]{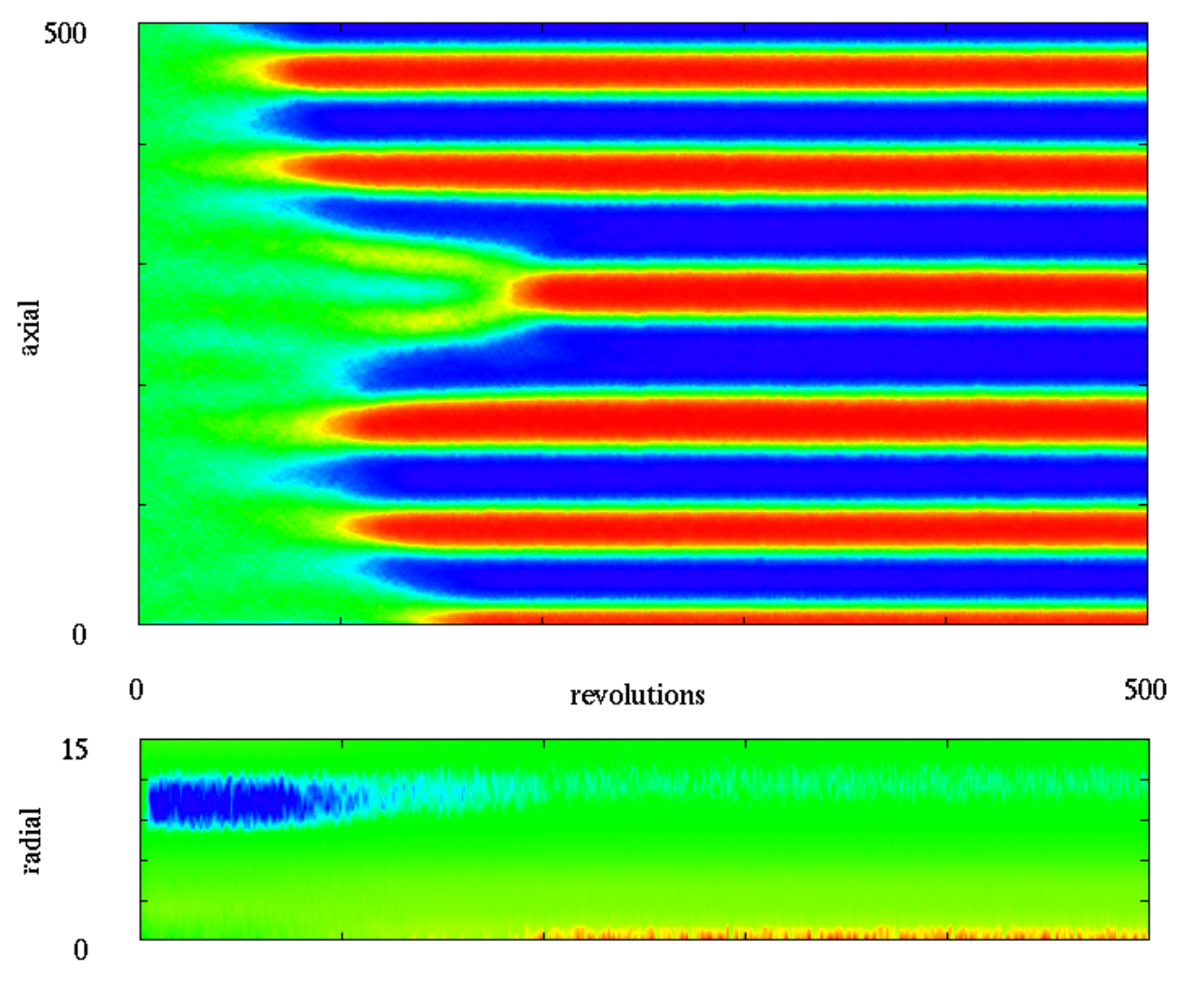}
\end{center}
\caption{\label{fig:grcyl_seg_plot}
Space-time plots showing axial and radial populations (longer cylinder); the
vertical axis corresponds to the axial or radial coordinates; red and blue
represent regions of predominantly big and small particles, while other
colors indicate mixed states.}
\end{figure}

Simulation of the longer cylinder leads to an entirely different mode of
behavior. The state after 1550 revolutions ($10^7$ timesteps) is shown in
Figure~\ref{fig:grcyl_ax_seg}; axial segregation is complete, with
relatively few particles appearing outside their segregation bands, as is
apparent from the views showing the individual species. During the early
phase of the run the system exhibits transient radial segregation, and the
space-time plots in Figure~\ref{fig:grcyl_seg_plot} show how radial and
axial segregation appear in succession; only the first 500 revolutions are
covered since no change occurs during the subsequent $> 1000$ revolutions.
Radial segregation appears rapidly, as before, but gradually fades after
about 100 revolutions, when axial segregation begins to develop. Once the
axial segregation process ends, there is very little particle transfer
between bands (as might be deduced from Figure~\ref{fig:grcyl_ax_seg}),
exactly as in \cite{rap07a}.

\subsection{Segregation in a vibrated layer}

Another instance of unusual behavior associated with granular matter
involves a vertically vibrated layer on a base whose surface is shaped as a
series of asymmetric sawtooth grooves. Horizontal flow is observed
\cite{far99}, both experimentally, where the grains are confined to a narrow
annular region between two upright cylinders, and in the corresponding
simulations \cite{lev01}. Moreover, the overall flow magnitude and direction
depend on the parameters defining the system, such as the vibration
frequency and amplitude, the sawtooth shape and the layer thickness, in a
seemingly unpredictable manner. The simulations reveal that the induced flow
rate actually varies with height within the layer; even more unexpected is
the observation that oppositely directed flows at different heights can
coexist, with flow close to the base directed up the less steep side of the
teeth. Once it is realized that the horizontal flow is stratified in this
fashion, the sensitive parameter dependence of the overall flow is seen to
be a direct consequence of the competition between opposing flows.

A combination of the stratified flow with another phenomenon unique to
granular matter, namely the familiar vertical size segregation under
vibration, or `Brazil nut' effect \cite{ros87,gal96}, leads to the
prediction of an entirely new kind of size-dependent segregation: Given that
under suitable conditions the upper and lower parts of the layer flow
horizontally in opposite directions when the sawtooth base is vibrated
vertically, and that these same vibrations cause vertical size segregation
within the layer, then it is reasonable to expect that a granular mixture
will undergo horizontal size segregation on a vertically vibrating sawtooth
base. Simulations \cite{rap01c} have demonstrated that this effect actually
occurs, at least in the computer.

Two examples of three-dimensional systems are considered, one a rectangular
box with linear sawtooth grooves, the other a cylindrical container with
concentric circular grooves; in \cite{rap01c} the two-dimensional version of
the box was considered, and the cylinder mentioned only briefly. The
granular model used here is simpler than for the rotating cylinder. The
grains are represented by soft spheres whose sizes are narrowly distributed
about two distinct mean values. The interactions between particles include
just the overlap repulsion (\ref{eq:ovint}) and normal damping
(\ref{eq:normdamp}), without the tangential component. Each sawtooth is
formed from several closely spaced, narrow, linear or ringlike cylinders,
horizontally and vertically positioned to approximate the profile; granular
particles near the sawtooth base interact with one or more of these
cylinders, with overlap repulsion and damping acting in the plane normal to
the (local) cylinder direction. Gravity is present, and the vertical
container walls are smooth.

The horizontal dimensions of the rectangular box are $270 \times 135$ while
the cylinder has a diameter of 270; in both cases the containers are
sufficiently high for the upper boundary to be beyond the range of particle
motion, and both contain $2 \times 10^5$ particles. The small particles have
unit maximum diameter, the big particles 1.5, with a small random value ($<
0.1$) subtracted to suppress any tendency to order; the fraction of big
particles is 0.3. The vibration frequency is $f = 0.4$, the amplitude $a =
1$, and $g = 5$; conversion to physical units is as before, resulting in a
frequency of 18\,Hz. The ratio of vibrational to gravitational acceleration,
$\Gamma = (2 \pi f)^2 a / g$, is a quantity relevant to the behavior of
vibrating layers, with a minimum $\Gamma \approx 1$ required to excite the
layer, and too large a value producing surface waves and eventually layer
breakup; here $\Gamma = 1.26$. Each gradually sloping sawtooth edge is of
approximate length 13, while the steep edge is practically vertical; the
sawtooth dimensions influence the effectiveness and even the direction of
the segregation process \cite{rap01c}.

\begin{figure}
\begin{center}
\includegraphics[scale=0.32]{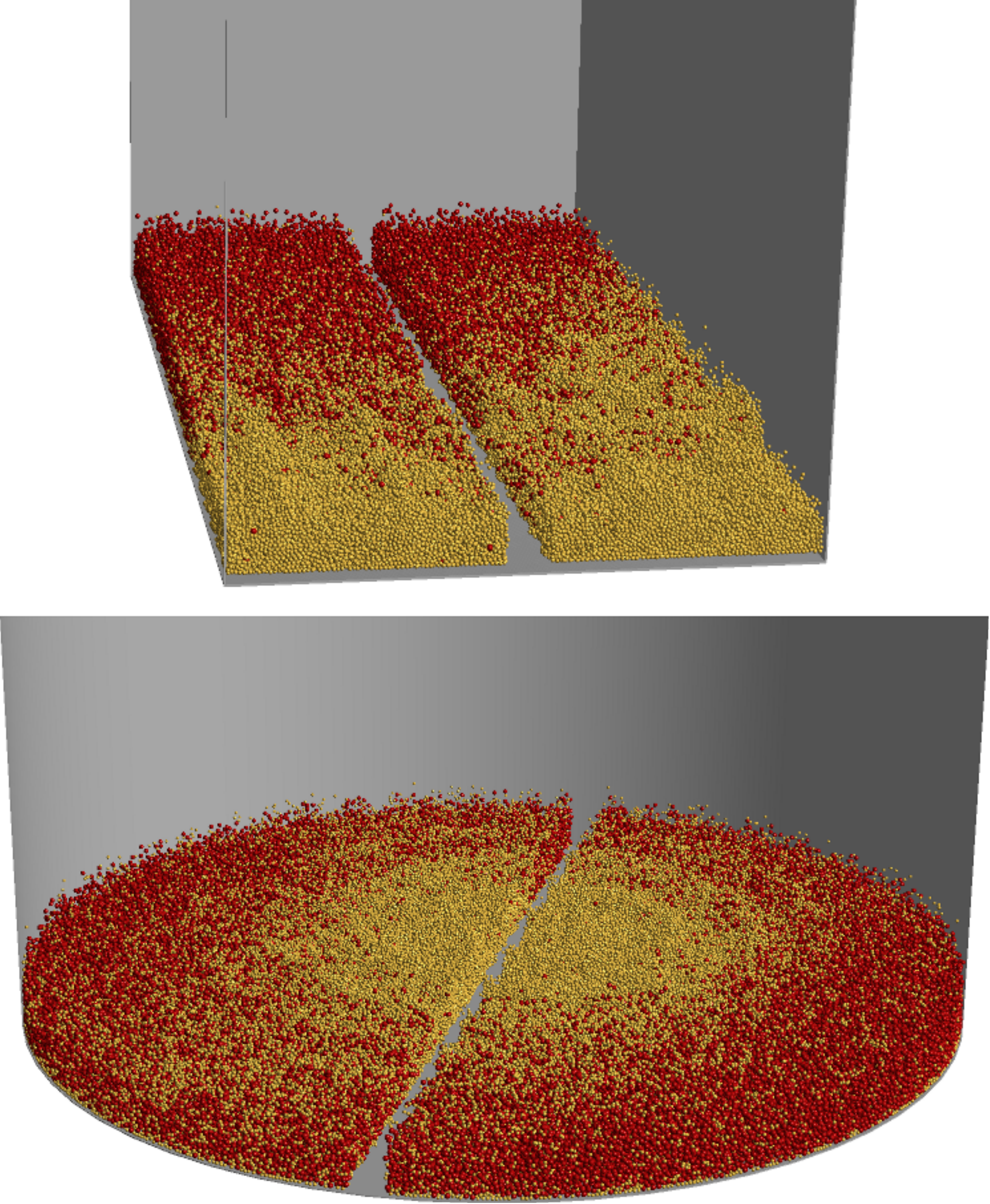}
\end{center}
\caption{\label{fig:grvib_both}
Horizontal segregation in vertically vibrated granular layers with a
sawtooth base; slices have been removed to shown the interior.}
\end{figure}

\begin{figure}
\begin{center}
\includegraphics[scale=0.22]{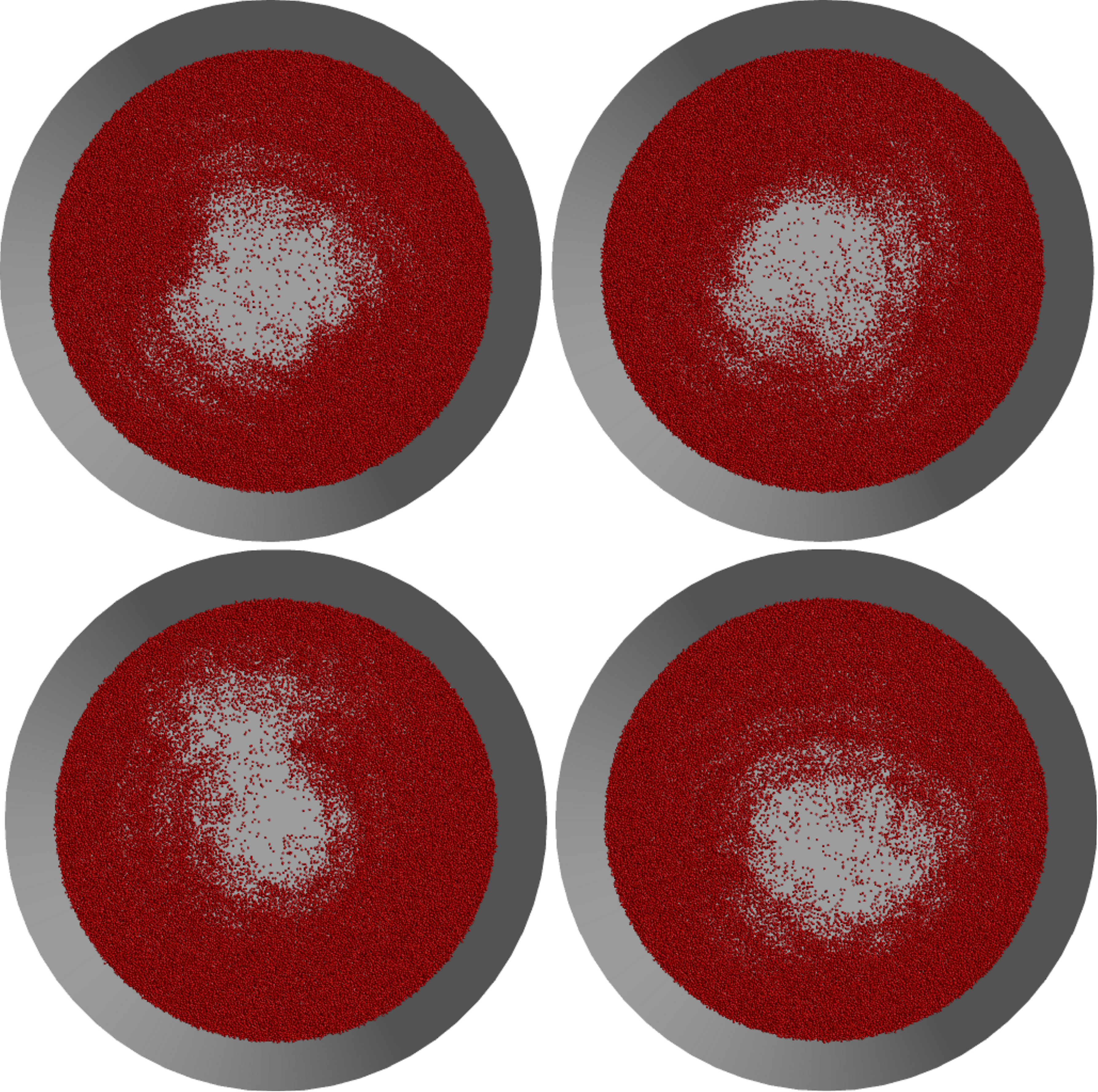}
\end{center}
\caption{\label{fig:grvib_cyl_seq}
Segregated big particles (viewed from above) at intervals of 1200 vibration
cycles.}
\end{figure}

The outcome of the simulations of the two systems is shown in
Figure~\ref{fig:grvib_both}, the box after $9 \times 10^4$ vibration cycles,
the cylinder after $4 \times 10^4$ cycles. Segregation requires only
$\approx 2000$ cycles to achieve its peak value, and the extended runs are
for monitoring long-term behavior. In each case the segregated state
persists, and the interface zone, where the species coexist, fluctuates
randomly. Snapshots showing variations in the region occupied by the big
particles (including the interface zone) for the case of the cylinder appear
in Figure~\ref{fig:grvib_cyl_seq}. The simulations are unequivocal in
demonstrating that this technique is able to produce segregation; it remains
to be seen whether there is agreement with experiment.

\section{Supramolecular self-assembly}

Supramolecular self-assembly underlies a broad range of emergent biological
phenomena at the molecular level; it is also the key to many proposed
industrial processes. One of the more fascinating examples of self-assembly
in nature is the spontaneous growth of the shells, or capsids, of spherical
viruses \cite{cri56,cas62} that package their genetic (RNA or DNA) payloads.
Capsids are assembled from multiple copies of one or a small number of
different capsomer proteins \cite{bak99} and exhibit icosahedral symmetry;
the structural organization is consequently simplified and the
specifications of the construction process minimized, important
considerations when all the details must be included in the genetic payload.

Since direct experimental observation of intermediate states along the
assembly pathway is difficult, little is known about the assembly process
itself, even for the simplified case where capsomers form complete shells
under {\em in vitro} conditions free of any genetic material
\cite{pre93,cas04,zlo11}. The overall robustness of the self-assembly
process \cite{cas80} justifies studying this simplified version of the
problem. There have been numerous simulational studies of self-assembly,
focusing both on capsid shells and other structures, with, e.g.,
\cite{hag06,fre06,ngu07} using MD and \cite{che07,wil07,joh10} employing
Monte Carlo methods that avoid dealing with the dynamics; a survey of capsid
assembly modeling appears in \cite{hag14}. Related `analog simulations' have
been performed in the laboratory using solutions of small plastic particles
with adhesive-coated surfaces \cite{bre99}. A considerable body of
theoretical work on capsid structure and assembly exists, based on a variety
of approaches including thin shells \cite{lid03}, particles on spheres
\cite{zan04}, tiling \cite{twa04}, stochastic kinetics \cite{hem06}, elastic
networks \cite{hic06}, nucleation theory \cite{zan06}, combinatorics
\cite{moi10}, and concentration kinetics \cite{mor09,hag10}, the last of
which is used for interpreting experiment \cite{zlo99,van07} and analyzing
reversible growth \cite{zlo07}.

The focus here is on the application of MD simulation to capsid growth using
particles whose shape is specifically designed to enable them to fit
together forming polyhedral shells \cite{rap99b,rap04a,rap08b,rap12c}. This
approach has proved successful, not only in achieving spontaneous
self-assembly of shells of different sizes, but also in allowing access to
the growth pathways and predicting how the populations of intermediate
structures vary over time, in principle providing a connection with
experiment \cite{end02}.

\begin{figure}
\begin{center}
\includegraphics[scale=0.24]{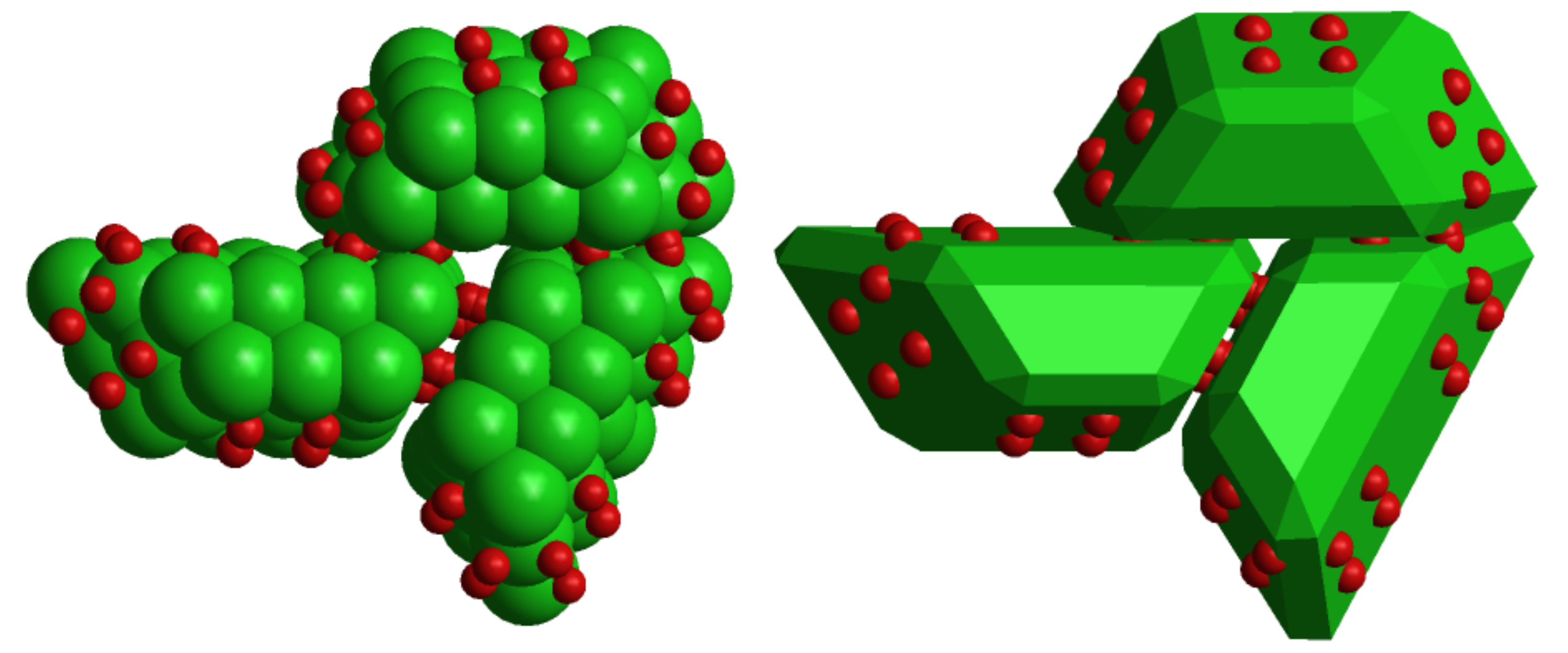}
\end{center}
\caption{\label{fig:sassem_trimer}
Fully-bonded trimer showing the spheres that define each particle, the
bonding sites (in red) and the effective particle shape.}
\end{figure}

Actual capsomers consist of large folded proteins whose exposed surfaces are
able to fit together to form strongly bound closed shells. On the other
hand, the design of simplified models for use with MD need only emphasize
two features of the capsomer, its overall shape and the attractive forces,
while avoiding the complexities of specific proteins. The present
simulations therefore deal with simple model particles, such as those in
Figure~\ref{fig:sassem_trimer}, that are constructed from sets of soft
spheres rigidly arranged to achieve the effective molecular shape required
for packing into a shell; to reproduce the organization of real capsids, the
particle shape is typically a truncated trapezoidal pyramid, in which all
edge lengths and facet angles are precisely determined. The figure shows
three particles in a fully-bonded trimer configuration; this structure is
not planar, unlike the other allowed fully-bonded trimer in which the
particles form a planar triangle.

The soft spheres on different particles interact via the usual short-range
repulsion (\ref{eq:ssint}). In order to define the bonding forces needed for
assembly, several interaction sites are located on the lateral faces of each
particle, shown in Figure~\ref{fig:sassem_trimer}, and attractive forces can
act between corresponding sites on specific face pairs of different
particles (consistent with the particle arrangement in the final shell
structure); the use of multiple interaction sites on each face ensures that
maximum attraction is achieved only when particles are properly aligned. The
interaction has the form of an inverse power law over most of its range, but
at small distances $< r_h$, it changes smoothly into a stretched harmonic
spring,
\begin{equation}
u_a(r) = \left\{
\begin{array}{ll}
e (1 / r_a^2 + r^2 / r_h^4 - 2 / r_h^2) & \quad r < r_h \\
e (1 / r_a^2 - 1 / r^2) & \quad r_h \le r < r_a
\end{array}
\right. \label{eq:attrint}
\end{equation}
where $e$ is the overall attraction strength that will be regarded as an
adjustable parameter. The range of the attractive force, $r_a = 3$, is
similar to the particle size (whose edge lengths vary between 2.1 and 3.6),
and the crossover distance is $r_h = 0.3$; this relatively narrow harmonic
well helps minimize structural fluctuations. While the attraction between
individual site pairs has no directional dependence, the involvement of
several pairs contributes to correct particle positioning and orientation,
further enhancing the rigidity of multiply-bonded structures. A side-effect
of the minimal structural fluctuations is that the final stages of shell
assembly are prolonged, because incoming particles must be correctly aligned
to fit into available openings that leave minimal room for maneuvering.

In order to carry out quantitative analysis of the cluster structure, those
particle pairs in which all four sets of matching attraction sites lie
within a prescribed (and arbitrary) range $r_b$ are considered bonded. There
is nothing special about bond formation since it is merely a bookkeeping
device, and the process is entirely reversible, a factor that turns out to
be central to successful assembly. A value of $r_b = 0.5$ ($> r_h$) leads to
results consistent with direct observation, namely that structural
fluctuations are responsible for only minimal spurious bond breakage and no
false bond designation. Each set of bonded particles forms a cluster. Since
the particle design and parameterization ensures that bonded pairs have very
limited relative motion, the only possible cluster in which every particle
has a full complement of five bonded neighbors is a closed shell of size 60;
mutant clusters do not develop for the range of $e$ considered here.

The simulations also include an explicit atomistic solvent, represented by
the same soft spheres (with repulsive interactions only) used in
constructing the particles. A thermostat is incorporated in the solvent
dynamics to regulate temperature \cite{rap04bk}; this is especially
important in view of the energy released as particles approach to within
bonding range. The explicit solvent entails significant computational cost,
but offers increased realism over the implicit (stochastic) alternative.
While both are capable of serving as heat baths to absorb the bond formation
energy, as well as curtailing the ballistic nature of the particle motion to
ensure conditions closer to thermal equilibrium, only the explicit approach
allows particles that have assembled into structures to offer mutual
shielding against disruptive solvent effects, aids cluster breakup without
subassemblies needing to collide directly, and incorporates the dynamical
correlations of the fluid medium. The choice of solvent representation is
also capable of affecting the outcome of self-assembly simulations
\cite{spa11}. The size ratio of the particles relative to the solvent atoms
is much smaller than in reality, in order to enhance particle mobility and
compress the time scales over which assembly occurs; the corresponding mass
ratio (here 15) is also reduced. Additional details appear in \cite{rap12c}.

Maximizing the yield of complete shells is an important goal in formulating
the model; the fact that the region size is limited makes this a more
important issue than it would be {\em in vivo} where other considerations
(of a more biological nature) are involved. Since allowing bond breakage
might be expected to reduce efficiency, the approach used both in the
original simulations \cite{rap99b} and as one of the alternatives in
\cite{rap04a} (see also \cite{rap10b}) is to make bond formation
irreversible (accomplished by altering the form of the pair attraction once
inside a suitably defined bonding range, together with a complicated
procedure aimed at avoiding bonds incompatible with the final structure). In
practice, it turns out that not only is reversible bonding much simpler from
a computational point of view (with incompatible bonds left to break on
their own) but, paradoxically, reversibility is a key contributor to
efficient assembly \cite{rap08b,rap12c}. Indeed, reversibility constitutes a
major difference between assembly at microscopic and macroscopic scales, and
is a consequence of the thermal `noise' that competes with the forces
driving growth, an effect practically invisible at the macroscopic level
(where Brownian motion offers a mere hint of its existence).

The large system described here has a total size of $N = 2.16 \times 10^5$,
almost twice that of \cite{rap12c} (where $1.25 \times 10^5$), to determine
the effect of size on the outcome. It includes $N_p = 4752$ particles,
enough for 79 complete shells of size 60, and a solvent with $N_a = N - N_p$
atoms. The particle fraction is the same as before, namely $p = N_p / N =
0.022$. The overall number density is 0.1; this determines the volume of the
cubic simulation region, and is a compromise that ensures the solvent can
fulfill its role while not excessively impeding particle motion. The
interaction strength parameter $e$ is set to 0.09, the value producing the
highest shell yield in the earlier work.

\begin{figure}
\begin{center}
\includegraphics[scale=0.44]{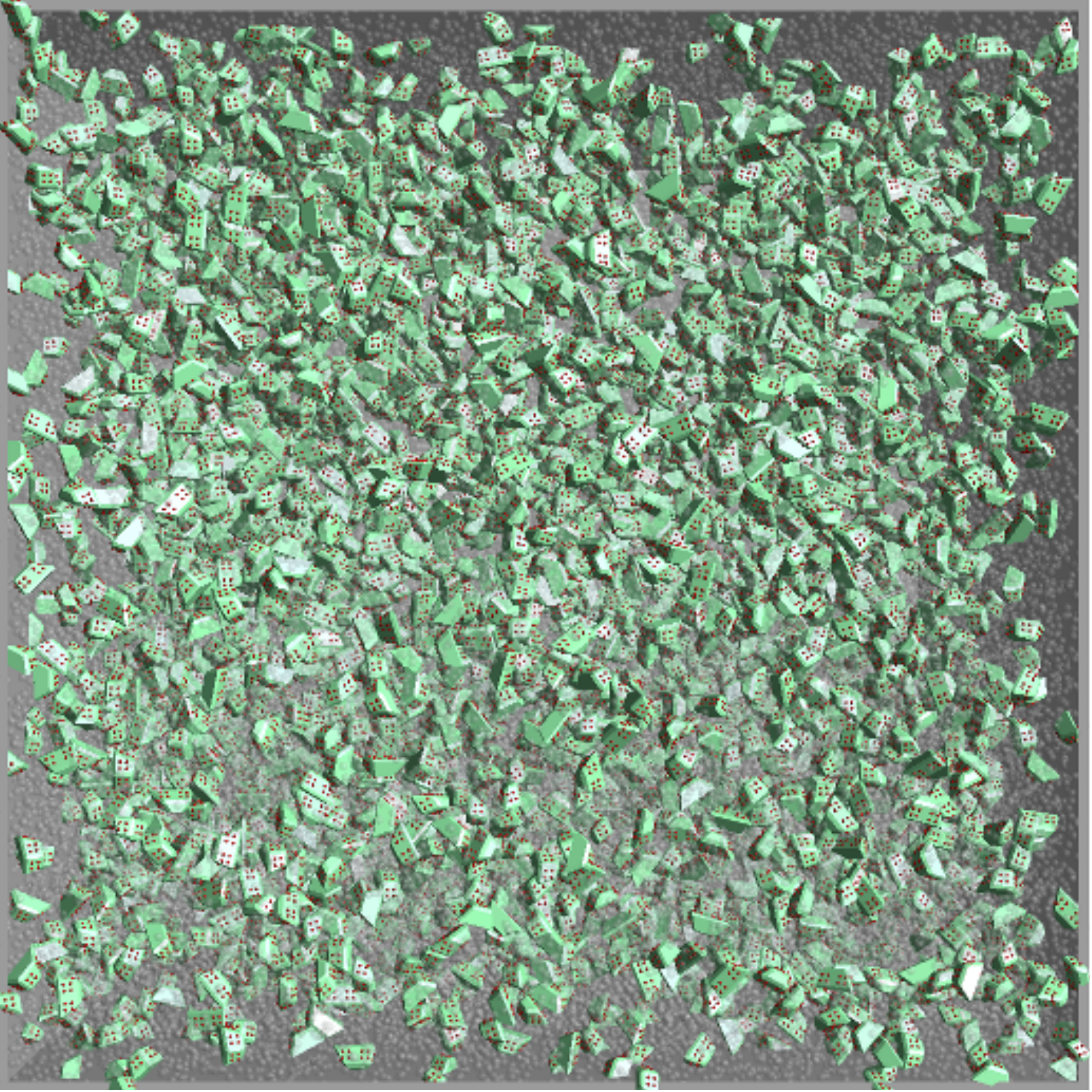}
\end{center}
\caption{\label{fig:sassem_big_bgn}
Early state of the large self-assembly simulation; the space-filling solvent
is shown semitransparently.}
\end{figure}

\begin{figure}
\begin{center}
\includegraphics[scale=0.44]{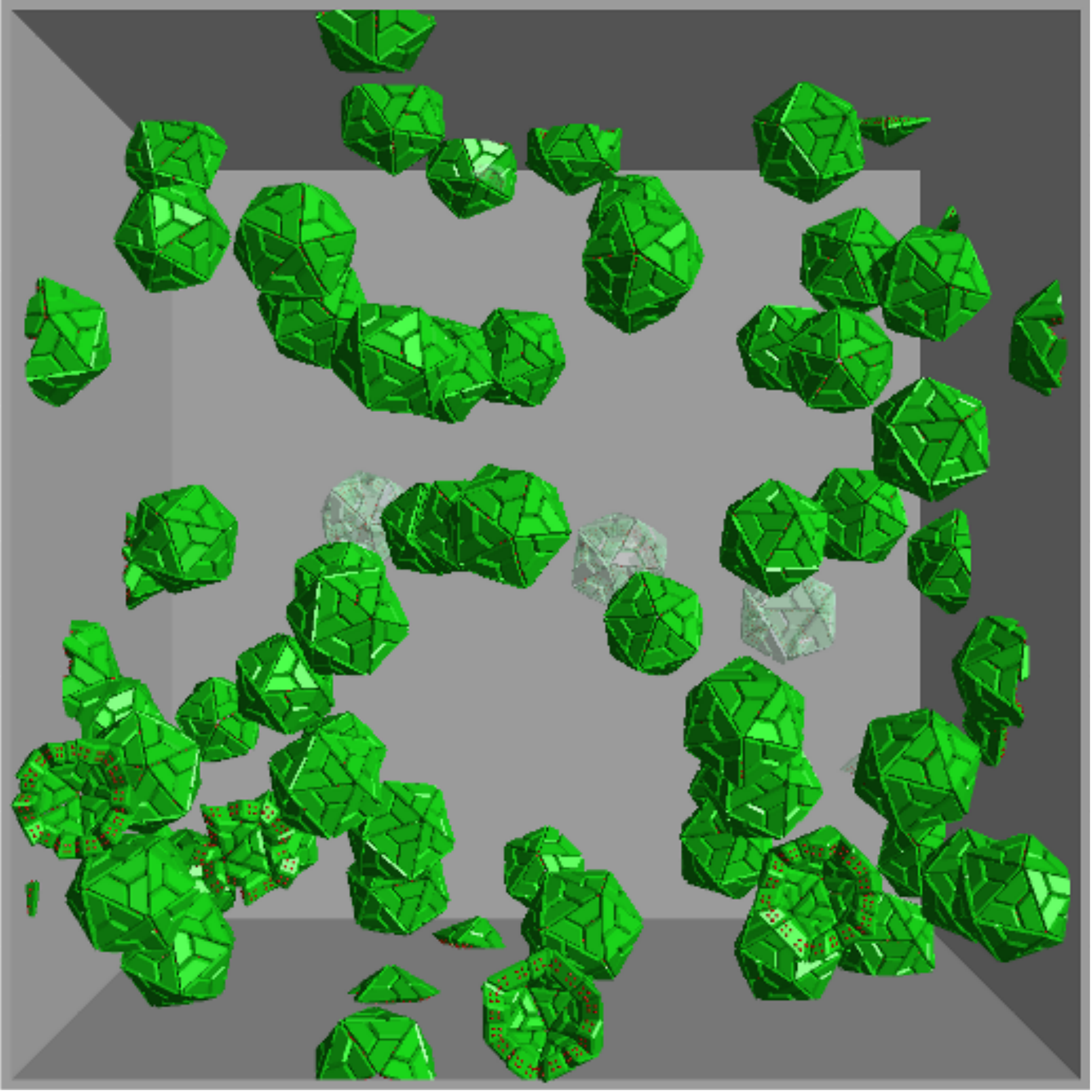}
\end{center}
\caption{\label{fig:sassem_big_end}
The final self-assembled state; monomers and solvent are omitted and
incomplete clusters are shown in a lighter color; because of periodic
boundaries some complete shells appear as fragments on opposite sides of the
region.}
\end{figure}

Figure~\ref{fig:sassem_big_bgn} shows the system at an early stage of the
run, after $5 \times 10^4$ timesteps, where only monomers are apparent. The
final state, after $3.2 \times 10^8$ timesteps, appears in
Figure~\ref{fig:sassem_big_end}; the remaining monomers and solvent are
omitted from the image so that the relevant particles can be seen. Here, the
57 complete shells that have formed can be seen (after allowing for the
periodic boundaries), a 72\% yield fraction, and just three incomplete
clusters. This picture of perfectly formed, self-assembled structures should
be contrasted with the previous image of an almost homogeneous early state 
(which, in fact, includes eight probably-transient dimers).

\begin{figure}
\begin{center}
\includegraphics[scale=0.56]{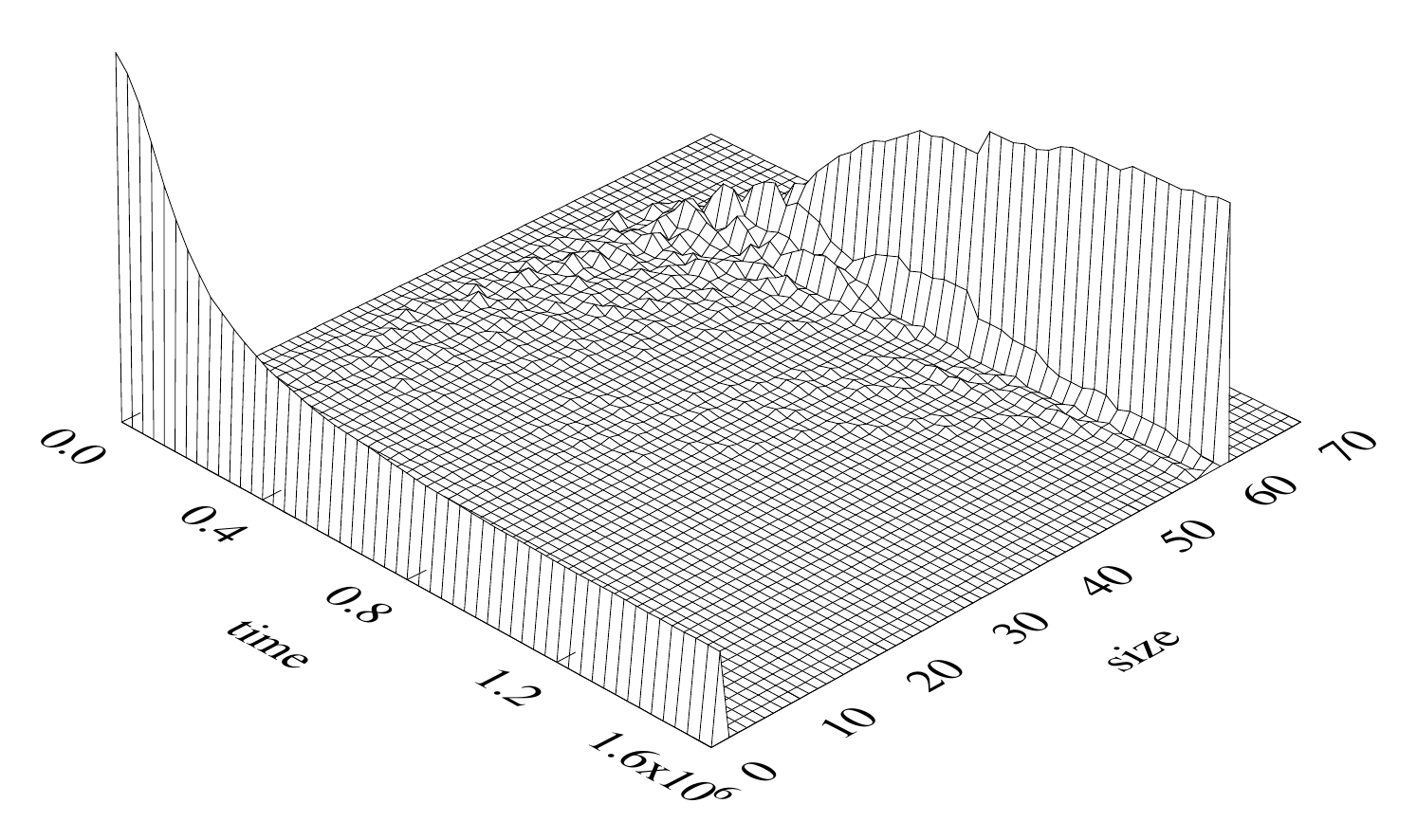}
\end{center}
\caption{\label{fig:sassem_big_mhist_plt}
The cluster mass fraction distribution for the large system as a function of
time; late in the run the system consists almost entirely of complete shells
and monomers.}
\end{figure}

The growth history of the shells, expressed in terms of the time- and
size-dependent distribution of the cluster mass fraction, is shown in
Figure~\ref{fig:sassem_big_mhist_plt}. The most notable feature of the graph
is the final state, which contains essentially only complete shells and a
significant residual monomer population; with the exception of just three
almost complete shells, intermediate-size clusters are entirely absent.
During the period over which most of the growth occurs the size distribution
is relatively broad and ill-defined, a consequence of the considerable
variation in the growth histories of individual clusters. Faint traces of
shells that are late developers are also apparent.

The lesson remains unchanged: given suitable parameters, it is possible to
achieve a final state consisting almost entirely of a high population of
complete shells and monomers. The successful self-assembly pathway, as
revealed by a more detailed analysis of cluster development \cite{rap12c},
comprises a cascade of reversible stages, with smaller clusters showing a
distinct preference for maximally bonded (i.e., low-energy) states,
eventually leading to a high yield of complete shells. This outcome is
attributable to the fact that reversibility maintains a low growth
initiation  probability, while at the same time allowing breakage of any
incorrect bonds that might attempt to form; alternative, less desirable
scenarios, such as minimal aggregation, multiple small clusters, and
clusters with unlimited growth, are all avoided. Reversibility dominates the
development of the smallest clusters, with dimers typically have just a 1\%
probability of growing into trimers (at $e = 0.09$), and trimers 50\% to
tetramers; over the broad size range 4--50 the probability of a unit size
increment is only $\approx 60\%$.

Once a closed shell has formed breakup is practically impossible. This is a
result of each particle being restrained by all five of its bonds, and the
absence of structural fluctuations capable of breaking individual bonds and
so initiate structural failure. This form of hysteresis was demonstrated
directly in \cite{rap08b}, by showing that if $e$ is reduced during the run
to a level at which assembly would not have occurred, all unfinished (and
therefore less strongly bound) assemblies disappear, leaving only the
complete shells.

\begin{figure}
\begin{center}
\includegraphics[scale=0.82]{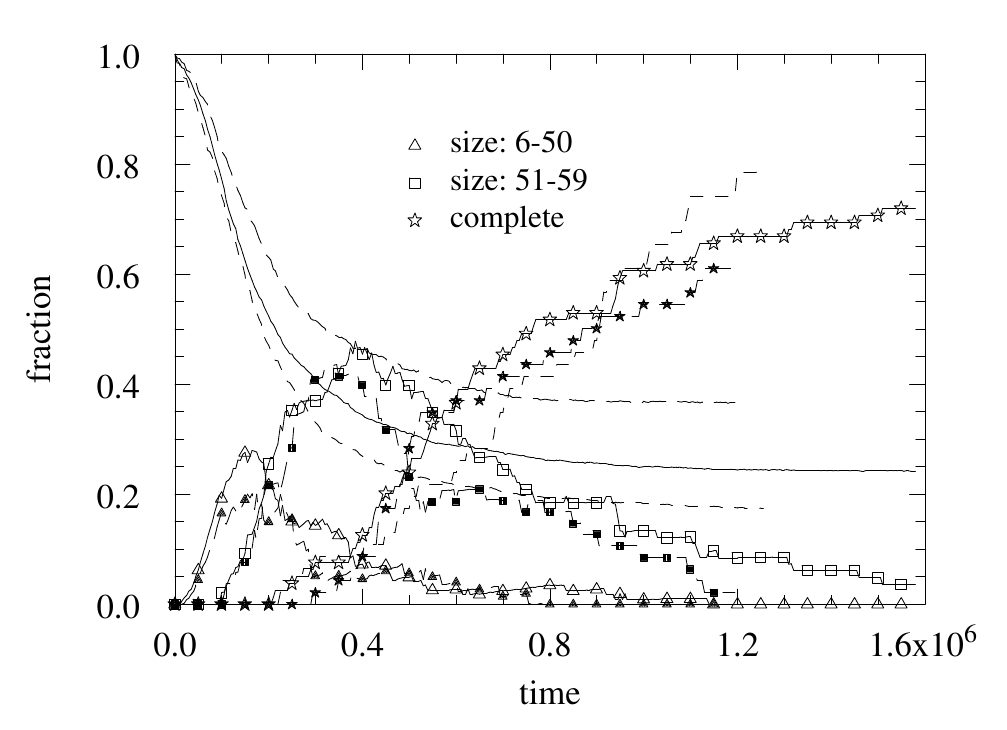}
\end{center}
\caption{\label{fig:sassem_cmp_mfrac_plt}
Time-dependent mass fractions for the large (solid curves, unfilled symbols)
and small (long dashes) systems, together with the earlier small system
\cite{rap12c} (short dashes, only monomers and complete shells shown);
results are grouped into size ranges, the monomer curves are without
symbols, and there is no contribution from the size range 2--5.}
\end{figure}

Multiple runs of identical systems starting from different initial states
reveal substantial variations in the number of complete shells. This is due
to the two rate-limiting growth stages, namely the low probability of
successfully initiating shell assembly by growing well beyond the dimer and
trimer clusters, and the prolonged delay for the final couple of assembly
steps to complete due to the difficulty of inserting the last monomers into
the small holes in the shell. Results from three runs are compared in
Figure~\ref{fig:sassem_cmp_mfrac_plt}. Two of the runs are for the smaller
system with different initial states, one of which is taken from
\cite{rap12c}. The differences between the complete-shell curves typify the
variability of the results; since the final state consists principally of
complete shells and monomers, the variations in the monomer mass fraction
mirror those of the complete shells. The third run is the large system,
where the fact that the fractional shell yield lies between the yields of
the two smaller systems shows that doubling the size leaves the behavior
unchanged. In view of the low survival probability of small clusters,
successful initiation of shell growth can be regarded as a rare event, with
consequences extending over the entire growth process and eventually
affecting the yield of complete shells. In this respect, the outcome
resembles the previous case studies, where effects amounting to little more
than noise are capable of altering the principal features of the final
result.

The final cluster distribution depends on $e$, with moderate to high yields
appearing only over a relatively narrow range; thus it is conceivable that
the high-yield phenomenon could easily have gone unnoticed. Results showing
the strong variation of the final distributions with $e$ for a 2.2\%
particle fraction, i.e., $p = 0.022$, were described in \cite{rap12c}.
Similar behavior occurs for other $p$; as an example, the final cluster
distributions of the smaller system ($N = 1.25 \times 10^5$) at $p = 0.033$
are tabulated in Table~\ref{tab:mass_frac} for various $e$. Allowing for the
inherent fluctuations, these results show the same kind of sensitive
$e$-dependence as for smaller $p$, ranging from no growth, through moderate
and high yields of complete shells, to a mixture of sizes but no complete
shells as $e$ is increased (note that raising $e$ is roughly equivalent to
reducing temperature).

When the particle fraction is lowered to $p = 0.015$, the shell yield is
greatly reduced for runs of length similar to those considered here,
although the possibility of continued slow growth cannot be excluded since
ample supplies of monomers remain. Given the increased elapsed time between
monomer encounters, and the short dimer lifetime, such a result is to be
expected, with the implication that a significant effort will be required
for computing the complete $p$\,--\,$e$ phase diagram. In view of the
protracted final stage of the assembly process, any comparison of the yields
for different $p$ and $e$ values, as required for the phase diagram
evaluation, might consider counting the nearly-complete shells together with
the complete shells, as well as taking into account whether the residual
monomer population is sufficient to fill the remaining holes given
sufficient time. A similar grouping of complete and nearly-complete shells
is relevant for comparison with experiment, since physical properties
dependent on cluster size might not be affected by a few small gaps in the
shells.

\begin{table}
\caption{\label{tab:mass_frac}
Final cluster mass fractions for different interaction strengths $e$,
grouped by cluster size (particle fraction $p = 0.033$).}
\begin{indented}
\item[]\begin{tabular}{@{}cccccc}
\br
$e$ & Timesteps & \multicolumn{4}{c}{Cluster mass fraction} \\
    & ($\times 10^6$) & Size: 1 & 2--50  & 51--59 & 60     \\
\mr
0.080 &  158 	      &  0.991  &  0.009 &  0	  &  0     \\
0.085 &  141 	      &  0.514  &  0.006 &  0.029 &  0.451 \\
0.088 &  171 	      &  0.278  &  0.010 &  0.014 &  0.698 \\
0.090 &  307 	      &  0.043  &  0.036 &  0.383 &  0.538 \\
0.095 & \ 85 	      &  0.018  &  0.479 &  0.503 &  0     \\
\br
\end{tabular}
\end{indented}
\end{table}

\section{Outlook}

A variety of systems exhibiting different forms of emergent behavior have
been studied using MD simulation. The phenomena associated with these
systems involve hydrodynamic instability, granular segregation and
supramolecular self-assembly. The most prominent feature in each of the
simulations is the emergence of a specific kind of ordered behavior
associated with, respectively, the flow patterns, spatial organization and
structure formation.

Although these problems have little in common, apart from the MD
methodology, since they deal with different systems and modes of behavior on
scales ranging from the molecular to the macroscopic, they share an ability
to produce effects that are both interesting and surprising. Given the
computationally-limited scope of MD simulation, the ability to achieve such
a rich set of outcomes is far from obvious. There are numerous reasons for
such skepticism, among them the possibility of capturing hydrodynamic
behavior within the relatively small spatial and temporal scales covered by
atomistic fluid simulation, the capability of simple frictional models to
replicate the complex behavior of granular mixtures (for one of the cases
considered, and predicting hitherto unknown behavior in the other), and the
ability of complex self-assembly processes to reach completion sufficiently
rapidly. The fact that each of the studies managed to generate meaningful
results is encouraging, as are the future prospects, given (still)
ever-increasing computational power and the availability of multiscale
methods (not covered here) to expand the simulational horizon.

An issue common to the problems considered here is reproducibility. The kind
of self-averaging that allows equating the results of a single MD simulation
to the ensemble average of statistical mechanics does not apply to the most
prominent results of the present case studies, namely the organization of
the induced flows, the spatial arrangement of the segregated regions, and
the yield of complete shells, nor to the intermediate stages of the
processes leading to these outcomes. Ensuring the reliability of any
conclusions based on these simulations requires repeating each of the runs
sufficiently many times to establish the `average' behavior and the range of
possible alternatives; this applies to each interesting parameter
combination, adding to the computational cost. It goes without saying that
the same lack of detailed reproducibility is intrinsic to the actual
physical systems that the simulations aim to emulate.

Finally, it is conceivable that there are some useful developments in
mathematics \cite{wig60} which might not have been forthcoming had `modern'
computers and their algorithms been available in an earlier era, especially
since the need to solve real physical problems was a motivating factor. The
absence of mathematical tools capable of addressing the inherent complexity
of emergent phenomena, despite ample motivation, guarantees the role of
computational exploration for the foreseeable future.

\section*{References}

\bibliographystyle{iopart-num}

\bibliography{emergephen}

\end{document}